\documentclass[preprint, 11pt]{elsarticle}
\usepackage{amssymb}
\usepackage{amsmath}
\usepackage{bm} 

\usepackage{lineno}

\usepackage{booktabs}

\usepackage{subcaption}

\usepackage{hyperref}
\hypersetup{
    colorlinks=true,
    linkcolor=blue,
    filecolor=magenta,      
    urlcolor=cyan,
    pdftitle={Effects of Horizontal Discretization}
    }

\usepackage{tikz} 
\usetikzlibrary {shapes.geometric}

\newcommand{\bsu}{\boldsymbol{u}} 
 
\newcommand{\bsb}{b} 
\newcommand{\bsp}{p} 
\newcommand{\bseta}{\eta} 
\newcommand{\bsw}{w} 
\newcommand{\abs}[1]{\lvert #1\rvert}

\DeclareMathSymbol{\shortminus}{\mathbin}{AMSa}{"39}

\DeclareRobustCommand{\redcircle}{\begin{tikzpicture}[baseline=-0.6mm]\node[shape=circle,fill=red,inner sep=1.5pt] {};\end{tikzpicture}}
\DeclareRobustCommand{\bluediamond}{\begin{tikzpicture}[baseline=-0.6mm]\node[shape=diamond,fill=blue,inner sep=1.5pt] {};\end{tikzpicture}}
\DeclareRobustCommand{\greensquare}{\begin{tikzpicture}[baseline=-0.6mm]\node[shape=rectangle,fill=green,inner sep=2pt] {};\end{tikzpicture}}

\begin{document}
\begin{frontmatter}

\title{Effects of Horizontal Discretization on Triangular and Hexagonal Grids on Linear Baroclinic and Symmetric Instabilities}

\author[awi]{Steffen Maaß}
\author[awi]{Sergey Danilov}

\affiliation[awi]{organization={Alfred Wegener Institute for Polar and Marine Research (AWI)},
            city={Bremerhaven},
            country={Germany}}

\begin{abstract}
As global ocean general circulation models are run at eddy-permitting
resolutions, reproducing accurate growth rates of baroclinic
instabilities is a major concern when choosing a discretization of the
equations of motion. From this viewpoint, we analyze discretizations on
triangular and hexagonal grids with different types of variable
staggering used in several ocean circulation models. By extending the
linear baroclinic instability analysis in the Eady configuration to discretizations on more complex grids, several
numerical subtleties are revealed.  In comparison to discretizations on
quadrilateral grids, the analyzed discretizations are less robust
against unstable spurious modes, partly created by the mesh geometry. Some of the subtleties arise because spurious modes on staggered triangular and hexagonal grids do not adhere to Galilean invariance. As a consequence, their growth rates demonstrate a dependence on the alignment between the background flow
and the grid, as well as the strength of a uniform background
flow. The interactions with spurious modes become more significant on
the axis of symmetric instabilities where the physical and spurious
branches of instability are more difficult to separate in
wavenumber space. Our analysis shows that in most cases moderate biharmonic viscosity and diffusion suppress spurious branches. However, one needs to carefully
calibrate the viscosity and diffusivity parameters for each of the
considered discretizations in order to achieve this.
\end{abstract}


\end{frontmatter}

\section{Introduction}
\label{sec:intro}
At present there are several global ocean general circulation models
formulated on unstructured meshes. They include MPAS-Ocean
(\cite{Ringler2013}), ICON-o (\cite{Korn2017a}) and FESOM
(\cite{Wang2014}, \cite{Danilov2017}). They use different types of
spatial discretization on horizontal planes.  The discretizations of
ICON-o and MPAS-Ocean are based on discrete normal velocities at the
edges of a primary triangular (ICON-o) or dual hexagonal mesh
(MPAS-Ocean) placing scalars at cells of the primary or dual mesh
respectively. FESOM works with full velocity vectors placed on cells
of a triangular mesh and scalars on median dual control volumes related
to mesh vertices. In analogy with the 'Arakawa' quadrilateral grids
(\cite{Arakawa1977}) the discretizations discussed here will be
referred to as the hexagonal C grid (MPAS-Ocean), triangular C grid
(ICON-o, even though this model uses a mimetic discretization
different from the usual C grid) and triangular B-grid (FESOM). The
list of unstructured-mesh ocean models and discretizations could have
been extended to include many coastal models but we do not consider
them here.

The question about advantages and disadvantages of regular
quadrilateral B and C grids has been discussed repeatedly and is
already a subject of textbooks (e.g. \cite{Haidvogel1999}). As is
well known, the B grids prevail in the representation of the Coriolis
force where the C grids introduce averaging but lose in the
representation of pressure gradients where they introduce additional
averaging compared to C grids. The recent analysis
\cite{Barham2018} adds to the understanding of the properties of the B
and C grids by studying how they reproduce known instabilities in the
Eady problem configuration. The resolution of global ocean circulation
models increases and numerous setups are used to simulate eddy-rich
regimes, and the analysis helps to assess efficient resolution of
these two grids (discretizations). \cite{Barham2018} shows that the C
grid discretization gives substantially more accurate growth rates of
symmetric instability but generates a spurious baroclinic (for
wavevectors aligned with the unperturbed flow) instability in the range of
wavenumbers extending up to the cutoff wavenumber. Suppression of the
latter requires a special combination of the fourth order scalar
advection and biharmonic horizontal viscosity. Thus, while both of the
grids are well suited to model the standard baroclinic instability,
the C grid can be a better option at very high resolutions when the
symmetric instability can become a part of dynamics.

The question arises as to whether these distinctions in reproducing
instabilities are preserved for triangular (hexagonal) C and B
grids. On one hand, the C grids still offer better locality of
pressure gradient computations than B grids. There are, however, many
differences compared to the case of quadrilateral grids. For example,
there is no pressure gradient averaging on triangular B grids, special
reconstructions are required for the Coriolis term on the hexagonal C
grid (\cite{Thuburn2009},\cite{Ringler2010}) and a special approach is
needed to express the horizontal divergence which is equivalent to
the averaging of primary normal velocities (\cite{Korn2017a}) on the
triangular C grid. There is an additional important difference,
however. All staggered unstructured grid discretizations maintain
spurious modes which emerge due to a disbalance between vector and
scalar degrees of freedom (DoF) (see, e.g., \cite{LeRoux2007, Thuburn2008, LeRoux2012, Danilov2013}). There are either too many scalar DoF
(on the triangular C grid) or too many velocity DoF (on the hexagonal
C grid and triangular B grid). Whether these spurious modes only modify the simulated growth rates or lead to spurious instabilities on their own is a question which remains to be answered.

The presented work is inspired by the analysis carried out in
\cite{Barham2018} and aims to extend this analysis to triangular and hexagonal mesh discretizations. We study the accuracy of these
discretizations in representing linear baroclinic and symmetric
instabilities in the Eady problem configuration. In analogy to
\cite{Barham2018}, the hydrostatic approximation is used together with the Boussinesq approximation and the buoyancy is considered as a single scalar field. 
Since the analysis is more cumbersome than on regular quadrilateral meshes, we present the cases of triangular B and C grids in the main text but computations of Fourier symbols are restricted to a few illustrative examples. Instead a notebook is provided in the Supplementary Material that allows the reader to further analyze the Fourier symbols of the linearized operators symbolically including operators on the hexagonal C grid. For completeness, and in contrast to staggered discretizations, the notebook also contains the Fourier symbols for the triangular A grid which, as expected, is much less prone to spurious instabilities.

\section{Methods}\setlength{\arraycolsep}{2pt}

The equations of motion are simplified by applying the Boussinesq and hydrostatic approximation. The momentum balance is assumed to hold on a $f$-plane with constant Coriolis parameter $f_0$ and biharmonic viscosity and diffusion with coefficients $\mathbb{K}^u$ and $\mathbb{K}^b$ respectively.
Different to \cite{Barham2018} we do not impose the rigid lid boundary condition at the surface but use a linearized free surface instead. In this case, a field of surface elevation, denoted by $\eta$, is introduced at scalar points and the reduced (divided by the reference density $\rho_0$) surface pressure becomes $g\eta$, where $g$ is the acceleration due to gravity. 
The limit of rigid lid is achieved taking the limit $g\to\infty$. 
In order to analyze the growth rates of the instabilities in the Eady configuration, the equations are linearized around a horizontal background flow 
\begin{equation*}
(U,V)=(\cos{\theta},\sin{\theta})M^2(z-H(1/2+\beta))/f_0, 
\end{equation*}
with the background buoyancy 
\begin{equation*}
B=N^2z-M^2x\sin{\theta}+M^2y\cos{\theta}.
\end{equation*}
Here $\theta$ is the angle of flow with respect to the $x$-axis and $\beta$ the non-dimensionalized speed of the background flow at mid-level. The parameters coincide with those in \cite{Barham2018}, i.e the flow depth is $H=4000$ m, the Coriolis parameter is $f_0=-10^{-4}$ s$^{-1}$, the Brunt-Väisälä (buoyancy) frequency is $N=10^{-3}$ s$^{-1}$. 
These choices yield a first defomation radius of order $R_d=NH/|f_0|=40$km (note the absence of $\pi$ in the denominator). The frequency, $M$,  of buoyancy in the horizontal direction is set by specifying the Richardson number $\mathrm{Ri}=N^2f_0^2/M^4$. 
In order to be comparable to \cite{Barham2018}, we consider the case $\mathrm{Ri}=100$ for baroclinic instability and $\mathrm{Ri}=0.5$ for the symmetric instability. The maximum growth rate for baroclinic instability is $0.31f_0/\sqrt{\mathrm{Ri}}$ at $kR_d\approx 1.6$.
The non-dimensional speed of the background flow at mid-level, $\beta$, is generally set to zero giving a profile symmetric around the mid-level but here the parameter is also varied to analyze the effect of the discretizations' violation of Galilean invariance.

The linearized equations of motions are summarized in the system \eqref{eq:linearized-eom} below.
The velocity field is split into a horizontal and a vertical component: $\mathbf{v} = (\mathbf{u}, w)$.
The system is closed with an impermeability condition at the bottom and a linearized free surface at the top.

\begin{subequations}
\begin{align}
\frac{\partial\bsu}{\partial t} + \left( \boldsymbol{U}(z) \cdot \nabla\right) \bsu + \boldsymbol{U}_z \bsw + f_0 \bsu^\perp + \mathbb{K}^u\Delta^2\bsu &= -\nabla\bsp  -g\nabla\bseta\\
\partial_z \bsp &= \bsb\label{eq:hydrostatic-eq}\\
\nabla\cdot \bsu + \partial_z \bsw &= 0\label{eq:continuity-eq}\\
\frac{\partial\bsb}{\partial t} + \left( \boldsymbol{U}(z) \cdot \nabla\right)\bsb + \nabla B\cdot \bsu + N^2 \bsw + \mathbb{K}^b\Delta^2\bsb &= 0 \\
\frac{\partial\bseta}{\partial t} + \left( \boldsymbol{U}(z) \cdot \nabla\right)\bseta + \int_{-H}^0 \nabla\cdot \bsu \mathrm{d}z &= 0 && \text{at}~~z=0\\
\bsp &= 0 && \text{at}~~z=0\label{eq:top-bc}\\
\bsw &= 0 && \text{at}~~z=-H\label{eq:bottom-bc}
\end{align}
\label{eq:linearized-eom}
\end{subequations}

The linearized system has constant coefficients on horizontal planes which lends itself to a Fourier transform of the horizontal coordinate space to wavenumber space. 
Using a layered-formulation of the equations of motion in the vertical direction and eliminating the vertical velocity and pressure by integration of the continuity equation \eqref{eq:continuity-eq} and the hydrostatic balance equation \eqref{eq:hydrostatic-eq}, a system of ODEs is obtained that can be solved for a fixed wavevector, $\boldsymbol{k}$.
Applying the usual ansatz in terms of the angular frequency, $\omega$, for the time dependence of the solutions to the corresponding eigenvalue problem, the solution is unstable if $\Im(\omega) > 0$ and stable otherwise.
The eigenvalue problem as formulated in equation \eqref{eq:eigenvalue-problem} is expressed in terms of the amplitudes of the horizontal velocities' plain waves, $\boldsymbol{\Hat{\underline{u}}} = (\underline{\Hat{u}}, \underline{\Hat{v}})$, the amplitudes of the buoyancy, $\underline{\Hat{b}}$, and the amplitude of the surface elevation, $\Hat{\eta}$, where the underlining denotes the vector of amplitudes on the different vertical layers.
The growth rates of the fastest growing instabilities on both the baroclinic and the symmetric axis are associated with the label ``ideal'' in Figure \ref{fig:bc} and \ref{fig:sy}.
\begin{equation}\label{eq:eigenvalue-problem}
    \frac{\mathrm{d}}{\mathrm{dt}}
    \renewcommand{\arraystretch}{1.3}
    \begin{bmatrix}
        \boldsymbol{\Hat{\underline{u}}} \\ \Hat{\underline{b}} \\ \Hat{\eta}
    \end{bmatrix}
    \exp(\mathit{i}(\mathbf{k}\cdot \mathbf{x} - \omega t))
    = \mathsf{S}_{\mathbf{k}} \begin{bmatrix}
        \boldsymbol{\Hat{\underline{u}}} \\ \Hat{\underline{b}} \\ \Hat{\eta}
    \end{bmatrix}
    \renewcommand{\arraystretch}{1.0}
    \exp(\mathit{i}(\mathbf{k}\cdot \mathbf{x} - \omega t))
\end{equation}

In order to analyze the effects of horizontal discretization on the instabilities' growth rates, the equations of motion are first formulated on the computational grid before the aforementioned procedure of linearization and Fourier transform is applied to obtain the discretized version of the eigenvalue problem \eqref{eq:eigenvalue-problem}.
One subtle difference is the number of amplitudes needed to specify the fields of the prognostic variables on a triangular (or hexagonal) mesh (see Table \ref{tab:amplitudes}).
This geometrical fact introduces the option of purely numerical branches of instability in the linearized problem.

Indeed, the geometry of triangular (or their dual hexagonal) meshes has certain specifics. In the language of crystallography, each primitive unit cell of a regular triangular lattice is composed of two triangles of different orientation. 
These triangles form two separate lattices which cannot be superimposed and hence should be distinguished in the analysis (see Figure \ref{fig:hex-lattice}). 
This distinction is the source of spurious/geometric modes that need to be dissipated
by appropriate viscosity and diffusion operators. 

We will be dealing with regular triangular meshes composed of equilateral triangles. The description of horizontal operators will be 
done in 2-dimensional geometry. While the considered discretization schemes are explained in detail for both triangular B grid (in section \ref{sec:method-B-type}) and triangular C grid (in section \ref{sec:method-C-type}) the reader interested only in one of the staggering types might skip to the corresponding section. 
A Lorenz staggering is applied for the discretization in the vertical direction.
Since the discussion is moslty the same as in the case of quadrilateral grids, the reader is referred to section 2 in \cite{Barham2018} for details.
Instead, in the following subsections the fields will be treated as continuous in the vertical direction and its discretization is only introduced in the summary of the system matrix in Equation \eqref{eq:TriB-sys-mat-alt} and briefly addressed in \ref{sec:appendixa}.
The growth rates of the fastest growing instabilities on a triangular grid as a function of wavenumber on the \textit{baroclinic} (\textit{symmetric}) axis are shown in Figure \ref{fig:bc} (resp. Figure \ref{fig:sy}). 

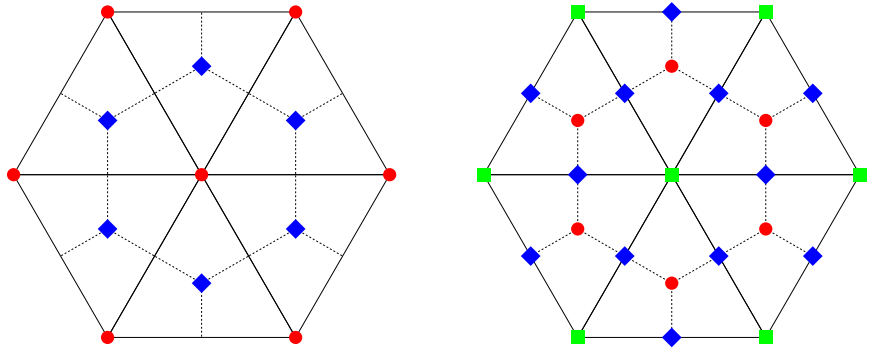
\begin{figure*}
    \centering
    \scalebox{0.25}{
    \begin{tikzpicture}
    \foreach \x/\y in {-15/0, -5/0, -10/{-sqrt(3)/2*10}}
    {
        \draw (\x+0,{\y+0}) -- ({\x+10},{\y+0});
        \draw ({\x+0},{\y+0}) -- ({\x+5},{\y+sqrt(3)/2*10});
        \draw ({\x+10},{\y+0}) -- ({\x+5}, {\y+sqrt(3)/2*10});
        \draw[dashed] ({\x+5},{\y+1/(sqrt(3)*2)*10}) -- ({\x+5},{\y});
        \draw[dashed] ({\x+5},{\y+1/(sqrt(3)*2)*10}) -- ({\x+2.5},{\y+ sqrt(3)/4*10});
        \draw[dashed] ({\x+5},{\y+1/(sqrt(3)*2)*10}) -- ({\x+7.5},{\y+ sqrt(3)/4*10});
    }
    \foreach \x/\y in {-15/0, -5/0, -10/{sqrt(3)/2*10}}
    {
        \draw (\x+0,{\y+0}) -- ({\x+10},{\y+0});
        \draw ({\x+0},{\y+0}) -- ({\x+5},{\y+ -sqrt(3)/2*10});
        \draw ({\x+10},{\y+0}) -- ({\x+5},{\y+ -sqrt(3)/2*10});
        \draw[dashed] ({\x+5},{\y-1/(sqrt(3)*2)*10}) -- ({\x+5},{\y});
        \draw[dashed] ({\x+5},{\y-1/(sqrt(3)*2)*10}) -- ({\x+2.5},{\y+ -sqrt(3)/4*10});
        \draw[dashed] ({\x+5},{\y-1/(sqrt(3)*2)*10}) -- ({\x+7.5},{\y+ -sqrt(3)/4*10});
    }

    \foreach \x/\y in {-15/0, -5/0, 5/0, -10/{sqrt(3)/2*10}, 0/{sqrt(3)/2*10}, -10/{-sqrt(3)/2*10}, 0/{-sqrt(3)/2*10}}
    {
        \node[circle, color=red, fill=red, minimum size=20pt] (v1) at ({\x},{\y}) {};
    }
    
    \foreach \x/\y in {-15/0, -5/0, -10/{-sqrt(3)/2*10}}
    {
        \node[diamond, color=blue, fill=blue, aspect=1, minimum size=30pt] (c1) at ({\x+5},{\y+1/(sqrt(3)*2)*10}) {};
    }

    \foreach \x/\y in {-15/0, -5/0, -10/{sqrt(3)/2*10}}
    {
        \node[diamond, color=blue, fill=blue, aspect=1, minimum size=30pt] (c1) at ({\x+5},{\y+ -1/(sqrt(3)*2)*10}) {};
    }

    \foreach \x/\y in {10/0, 20/0, 15/{-sqrt(3)/2*10}}
    {
        \draw (\x+0,{\y+0}) -- ({\x+10},{\y+0});
        \draw ({\x+0},{\y+0}) -- ({\x+5},{\y+sqrt(3)/2*10});
        \draw ({\x+10},{\y+0}) -- ({\x+5}, {\y+sqrt(3)/2*10});
        \draw[dashed] ({\x+5},{\y+1/(sqrt(3)*2)*10}) -- ({\x+5},{\y});
        \draw[dashed] ({\x+5},{\y+1/(sqrt(3)*2)*10}) -- ({\x+2.5},{\y+ sqrt(3)/4*10});
        \draw[dashed] ({\x+5},{\y+1/(sqrt(3)*2)*10}) -- ({\x+7.5},{\y+ sqrt(3)/4*10});
    }
    \foreach \x/\y in {10/0, 20/0, 15/{sqrt(3)/2*10}}
    {
        \draw (\x+0,{\y+0}) -- ({\x+10},{\y+0});
        \draw ({\x+0},{\y+0}) -- ({\x+5},{\y+ -sqrt(3)/2*10});
        \draw ({\x+10},{\y+0}) -- ({\x+5},{\y+ -sqrt(3)/2*10});
        \draw[dashed] ({\x+5},{\y-1/(sqrt(3)*2)*10}) -- ({\x+5},{\y});
        \draw[dashed] ({\x+5},{\y-1/(sqrt(3)*2)*10}) -- ({\x+2.5},{\y+ -sqrt(3)/4*10});
        \draw[dashed] ({\x+5},{\y-1/(sqrt(3)*2)*10}) -- ({\x+7.5},{\y+ -sqrt(3)/4*10});
    }

    \foreach \x/\y in {10/0, 20/0, 30/0, 15/{sqrt(3)/2*10}, 25/{sqrt(3)/2*10}, 15/{-sqrt(3)/2*10}, 25/{-sqrt(3)/2*10}}
    {
        \node[rectangle, color=green, fill=green, minimum height=20pt, minimum width=20pt] (v1) at ({\x},{\y}) {};
    }
    
    \foreach \x/\y in {10/0, 20/0, 15/{-sqrt(3)/2*10}}
    {
        \node[circle, color=red, fill=red, minimum size=20pt] (c1) at ({\x+5},{\y+1/(sqrt(3)*2)*10}) {};
    }

    \foreach \x/\y in {10/0, 20/0, 15/{sqrt(3)/2*10}}
    {
        \node[circle, color=red, fill=red, minimum size=20pt] (c2) at ({\x+5},{\y+ -1/(sqrt(3)*2)*10}) {};
    }

    \foreach \x/\y in {10/0, 20/0, 15/{sqrt(3)/2*10}, 15/{-sqrt(3)/2*10}}
    {
        \node[diamond, color=blue, fill=blue, minimum size=30pt] (e3) at ({\x+5}, {\y+0}) {};
    }

    \foreach \x/\y in {10/0, 20/0, 15/{sqrt(3)/2*10}, 25/{sqrt(3)/2*10}}
    {
        \node[diamond, color=blue, fill=blue, minimum size=30pt] (e1) at ({\x+2.5}, {\y+ -sqrt(3)/4*10}) {};
    }

    \foreach \x/\y in {10/0, 20/0, 15/{-sqrt(3)/2*10}, 25/{-sqrt(3)/2*10}}
    {
        \node[diamond, color=blue, fill=blue, minimum size=30pt] (e2) at ({\x+2.5}, {\y+sqrt(3)/4*10}) {};
    }
    
    \end{tikzpicture}
}
    \caption{Variable staggering on triangular B (left panel) and C (right panel) grids. Scalar quantities are located at the red circles (``\redcircle'') and their values are interpreted as mean values of the corresponding control volume (\textit{left}: median-dual control volume with hexagonal boundary of dotted lines obtained by connecting cell centers with mid-edges, \textit{right}: cell with triangular boundary of solid lines) in a finite volume formulation. The vectors of horizontal velocities on B grids and normal velocities on C grids are at the blue diamonds (``\bluediamond''). On B grids the their values are interpreted as mean values of the corresponding triangular cell. On C grids the (vertical) vorticity is staggered with respect to the scalar quantities and its values are located at the green squares (``\greensquare'').}
    \label{fig:staggering}
\end{figure*}

\subsection{Triangular Grid with B-type Staggering}\label{sec:method-B-type}
On a triangular B grid the discrete horizontal velocities are
located at the centroids of the triangles. In a finite volume
formulation their values may be interpreted as the mean on the
respective triangles which are called \textit{cells} in this
context. The scalar quantities such as surface height, pressure,
buoyancy and vertical velocity are staggered with respect to the
horizontal velocities and are collocated at the vertices of the
grid. Their values may be interpreted as mean values on the median-dual control volumes. The latter are obtained by connecting cell centers with mid-edges. For a regular triangular mesh the median-dual control volumes coincide with the hexagonal cells of the dual mesh, and we will not distinguish between them further in the text. See the left panel of Figure \ref{fig:staggering} for an illustration of the B-type staggering on a triangular mesh.

\begin{table}
\centering
\begin{tabular}{ll}
\toprule
Notation & Interpretation \\[0pt]
\midrule
$c$ & cell\\[0pt]
$v$ & dual control volume\\[0pt]
$e$ & face (edge) between control volumes\\[0pt]
\(\boldsymbol{n}_{c,e}\) & outer normal to cell c at face e\\[0pt]
\(\boldsymbol{n}_{v,e}\) & outer normal to dual control volume v at face e\\[0pt]
\(\boldsymbol{n}_{e}\) & orientation of the face (edge) e\\[0pt]
\(\boldsymbol{t}_e\) & orientation of the dual face (edge) \(\Hat{e}\)\\[0pt]
\(\ell_e\) & area (length) of the face (edge) e\\[0pt]
\(d_e\) & area (length) of the dual face (edge) e\\[0pt]
\(E(c)\) & set of faces of cell c\\[0pt]
\(E(v)\) & set of faces of dual control volume v\\[0pt]
\(C(e)\) & set of cells with face e\\[0pt]
\(V(e)\) & set of dual control volumes with face e\\[0pt]
\bottomrule
\end{tabular}
\caption{Notations for the discretization on a triangular grid.}
\label{tab:notation-Tri}
\end{table}

The horizontal discretization of the governing equations may be
formulated in terms of discrete analogs of the differential operators.
The discrete operators may be interpreted as (the sum of pointwise
products of) discrete convolutions on these lattices. The kernels may
be defined on a lattice isomorphic to the lattice of the respective
input signal but shifted to a collocation point of the respective
output signal. The Fourier symbol of a discrete operator corresponding
to a constant coefficient convolution kernel is then given by the discrete Fourier transform of the respective convolution kernels.

\subsubsection{Discrete Linear Operators}
Using the notation of \cite{Danilov2017} (see Table \ref{tab:notation-Tri})  discrete
analogs of differential operators may be formulated on the B grid. Here, the same notation as for their continuous counterparts is used for fields defined on the discrete horizontal spaces.

\begin{subequations}
\begin{align}
\big( \nabla \bsp\big)_c
&=\frac{1}{A_c} \sum_{e\in E(c)} \boldsymbol{n}_{c,e}\ell_e \frac{1}{2}\sum_{v\in V(e)} p_v \label{eq:TriB-gradient-scalar}\\
\big( \nabla \boldsymbol{F}\big)_c
&=\sum_{e\in E(c)} \frac{2}{3} \boldsymbol{n}_{c,e} \frac{1}{d_e}\sum_{\Tilde{c}\in C(e)} \boldsymbol{n}_{c,e}^T\boldsymbol{n}_{\Tilde{c},e}  \boldsymbol{F}_v^T
\label{eq:TriB-gradient-vector}\\
\big(\nabla\cdot\boldsymbol{F}\big)_v
&= \frac{1}{A_v} \sum_{e\in E(v)}d_e\frac{1}{2} \sum_{c\in C(e)} \boldsymbol{n}_{v,e} \cdot \boldsymbol{F}_c
\label{eq:TriB-divergence}\\
\Big( \mathrm{Curl}~\bsu\Big)_v 
&= \sum_{e\in E(v)} \sum_{c\in C(e)} \frac{\ell_e}{2} \boldsymbol{n}_{v,e}^\perp \cdot \bsu_c
\label{eq:TriB-curl}\\
\Big(\Delta \bsb \Big)_v
&= \Big(\nabla \cdot \nabla \bsb\Big)_v
\label{eq:TriB-laplacian}\\
\Big(\Delta \bsu \Big)_c
&= \frac{1}{A_c}\sum_{e\in E(c)} \ell_e \frac{1}{d_e}\sum_{\Tilde{c}\in C(e)} -\boldsymbol{n}_{c,e}^T\boldsymbol{n}_{\Tilde{c},e} \bsu_{\Tilde{c}}
\label{eq:TriB-vector-laplacian}
\end{align}
\end{subequations}

The discrete gradient operator acting on the pressure field averages
the pressure values along edges and identifies the gradient with the
normal to the unique plane defined by the pressure averages on the
three boundary edges of the respective cell (Equation \eqref{eq:TriB-gradient-scalar}).
If the pressure field is a plain wave with wavevector $\boldsymbol{k}$ then the gradient operator may be expressed in terms of a matrix acting on the amplitude of the plain wave, $\Hat{p}(\boldsymbol{k})$.
Here the matrix has size $4\times 1$ because on a triangular grid with B-type staggering a pressure field requires only one amplitude whereas the resulting gradient is a vector quantity defined on cells and requires four amplitudes (for both the $x$- and $y$-direction one amplitude on both the lattice of upward-pointing cells and the lattice of downward-pointing cells).
The matrix is called is called the \textit{Fourier symbol} of the gradient operator\footnote{Here and thereafter in most cases the explicit dependence of the Fourier symbol on the wavevector is neglected in the notation.}:
\begin{equation*}
\mathsf{G} = \begin{bmatrix} g_x^u &
g_x^d & g_y^u & g_y^d\end{bmatrix}^T
\end{equation*}
where \(g_x\) and \(g_y\) are the
components of the gradient and the superscript indicates the lattice
of upward-pointing and downward-pointing cells respectively.
The components satisfy the following relations:

\begin{align*}
    \big( \nabla \bsp\big)_{c^u} &= e^{i\boldsymbol{k}\cdot \boldsymbol{x}_{c^u}} \begin{bmatrix}g_x^u(\boldsymbol{k}) & g_y^u(\boldsymbol{k}) \end{bmatrix}^T  \Hat{p}(\boldsymbol{k})\\
    \big( \nabla \bsp\big)_{c^d} &= e^{i\boldsymbol{k}\cdot \boldsymbol{x}_{c^d}} \begin{bmatrix}g_x^d(\boldsymbol{k}) & g_y^d(\boldsymbol{k}) \end{bmatrix}^T \Hat{p}(\boldsymbol{k})\\
    &= e^{i\boldsymbol{k}\cdot \boldsymbol{x}_{c^d}} \begin{bmatrix}\shortminus\left(g_x^u(\boldsymbol{k}) \right)^* & \shortminus\left(g_y^u(\boldsymbol{k}) \right)^*\end{bmatrix}^T \Hat{p}(\boldsymbol{k}).
\end{align*}

As an example the components of the gradient's Fourier symbol are explicitly computed for upward-pointing cells.
The notations of the relative phases are given in right panel of Figure \ref{fig:hex-lattice}.

\begin{equation*}
\begin{split}
\big( \nabla \bsp\big)_{c^u}
&= e^{i\boldsymbol{k}\cdot \boldsymbol{x}_{c^u}} \frac{a}{2A_c} \bigg( \begin{bmatrix}0\\-1\end{bmatrix} \left(e^{-i\alpha_1} + e^{-i\alpha_2}\right)+ \begin{bmatrix}\sqrt{3}/2\\1/2\end{bmatrix} \left(e^{-i\alpha_2}+ e^{-i\alpha_3}\right)\\
&\qquad+ \begin{bmatrix}-\sqrt{3}/2\\1/2\end{bmatrix}\left(e^{-i\alpha_3} + e^{-i\alpha_1}\right)  \bigg) \Hat{p}(\boldsymbol{k})\\
&= e^{i\boldsymbol{k}\cdot \boldsymbol{x}_{c^u}}\begin{bmatrix}
\left( e^{-i\alpha_2} - e^{-i\alpha_1} \right)/a\\ \left( 2e^{-i\alpha_3} - e^{-i\alpha_1} - e^{-i\alpha_2} \right)/(2h)
\end{bmatrix}  \Hat{p}(\boldsymbol{k}).
\end{split}
\end{equation*}

On a regular grid the discrete gradient operator acting on a vector
field can be expressed by Equation \eqref{eq:TriB-gradient-vector} and it corresponds to the
more general least square formulation on unstructured grids
(cf. \cite{Danilov2017}). Its Fourier symbol in the $x$-direction can be
expressed in terms of the previously discussed gradient's Fourier
symbol components and its skew-hermitian property should be noted.

The discrete divergence operator \eqref{eq:TriB-divergence} acting on a flux field
averages the normal fluxes along edges to compute the total flux
through the six boundary edges of the scalar control volume, \(v\). The
Fourier symbol of the discrete divergence operator is related to the
negative Hermitian transpose of the gradient's Fourier symbol: $\mathsf{D} = -\mathsf{G}^*/2$.

By Stokes's theorem the vertical component of the vorticity on a scalar control volume, $v$, may be expressed in terms of the line integral of the horizontal velocity field along the boundary of the control volume.
The discrete curl operator \eqref{eq:TriB-curl} acting on a velocity field uses a simple approximation of the line integral along the boundary.
Its Fourier symbol is not explicitly given here but instead the interested reader can analyze it using the notebook provided as Supplementary Material.

The discrete Laplace operators on scalar control volumes \eqref{eq:TriB-laplacian} is simply the divergence of the scalar field's gradient.
Thus, its Fourier symbol is the product of the Fourier symbols of the divergence and gradient operators:
\begin{equation}
    \mathsf{L^b} = \mathsf{D}\mathsf{G} = -\frac{1}{2} \mathsf{G^* G}.
\end{equation}
In particular, the Fourier symbol is a negative real number and the symbol for biharmonic diffusion, $\mathsf{D^b = -\mathbb{K}^b \mathsf{L}^b\mathsf{L}^b}$, is also negative for positive diffusion coefficient, $\mathbb{K}^b >0$. 

The Fourier symbol, in the following denoted by $\mathsf{L^u}$, of the discrete Laplace operator on cells acting on a vector field \eqref{eq:TriB-vector-laplacian} is block-diagonal with $2\times 2$ Hermitian negative definite block matrices but not diagonal.
The symbol for biharmonic viscosity is the negative product of the viscosity coefficient, $\mathbb{K}^u$, with the square of the Laplace operator's symbol: $\mathsf{D^u}=-\mathbb{K}^u\mathsf{L^u}\mathsf{L^u}$.
As a consequence the symbol is also Hermitian negative definite for positive viscosity parameters but in general not diagonal.
In particular, a change in the viscosity parameter may result in a change of the eigenmodes of the system matrix.
Again the interested reader is referred to the accompanying notebook for more details.

\subsubsection{Transport of Horizontal Momentum}
\begin{figure}
    \begin{subfigure}{0.5\linewidth}
        \centering
        \includegraphics[width=0.99\linewidth]{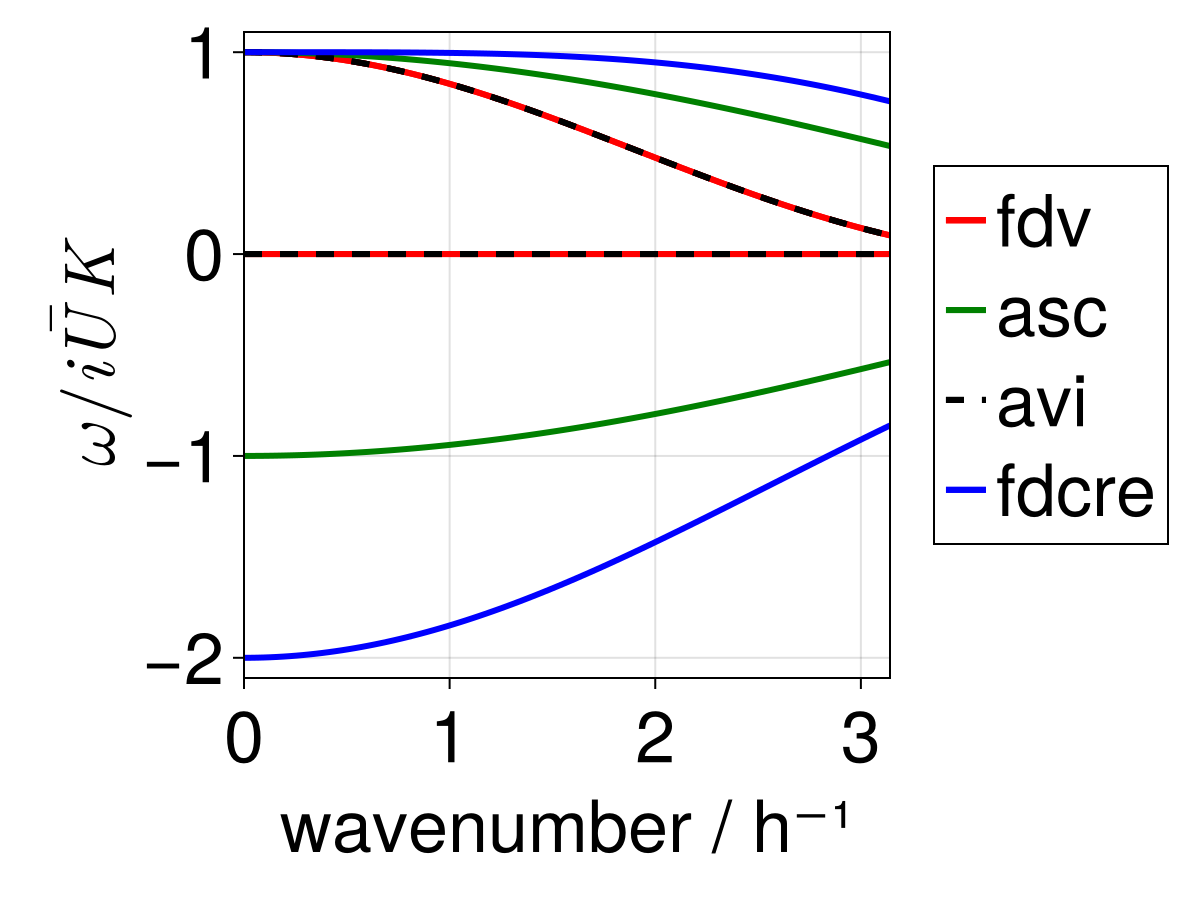}
        \phantomcaption
         \label{fig:hmt-FS}
    \end{subfigure}
    \begin{subfigure}{0.49\linewidth}
        \centering
        \includegraphics[width=0.99\linewidth]{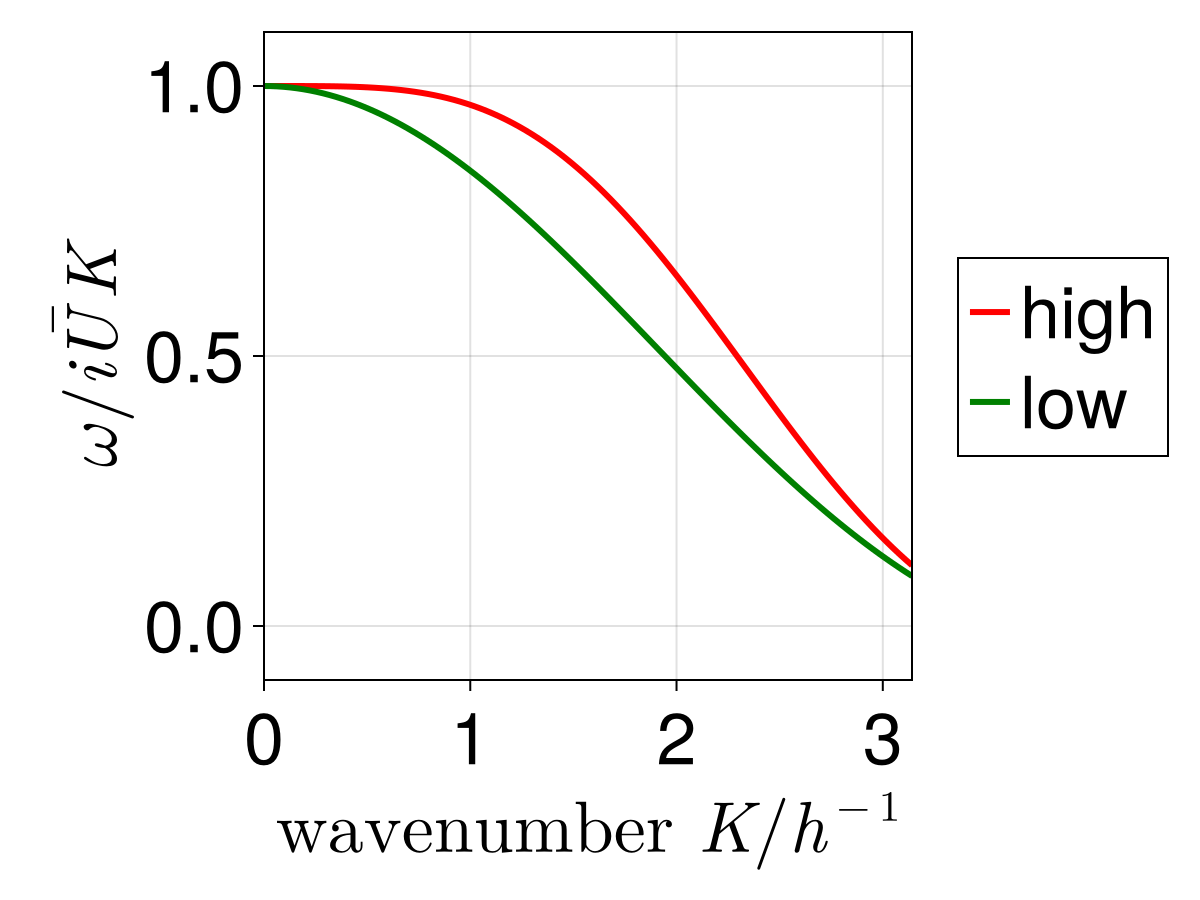}
        \phantomcaption
        \label{fig:hstb-FS}
    \end{subfigure}
    \caption{\textit{Left}: Eigenvalues of Fourier symbols for different linearized momentum advection schemes on triangular B grid as function of the wavenumber in the flow direction. The schemes $\mathrm{FDCRE}$ and $\mathrm{ASC}$ have more accurate physical branches but in addition a spurious/geometric branch of waves traveling in the direction opposite to the flow direction. The eigenvalues of the schemes $\mathrm{FDV}$ and $\mathrm{AVI}$ coincide, and the spurious branch has zero frequency, but note that the spurious branches in the Eady configuration with the respective schemes show different behavior. \textit{Right}: Fourier Symbol of the linearized advection schemes acting on buoyancy as a function of the wavenumber in the flow direction. The advection schemes are defined on a triangular grid with B-type staggering and linearized around a constant flow.}\label{fig:TriB-adv-ops}
\end{figure}

The transport of horizontal momentum may be discretized in the advective form, flux form
or the vector invariant form.  Here the analysis is restricted to four
different schemes. 
In analogy to the linearized advection operator in the continuous space (see Equation \eqref{eq:linearized-eom}) the Fourier symbol of the linearized advection operator in the discrete space is written in the form $U\mathsf{G}_x(\boldsymbol{k}) + V\mathsf{G}_y(\boldsymbol{k})$ and acting on the four amplitudes that specify the horizontal velocity field as a plane wave with fixed wavevector $\boldsymbol{k}$.
The matrices $\mathsf{G}_x$ and $\mathsf{G}_y$ are of size $4\times 4$ because on a triangular grid with B-type staggering the horizontal velocity field is a vector quantity defined on cells.
Figure \ref{fig:hmt-FS} shows the eigenvalues of the
corresponding Fourier symbols of linearized momentum advection evaluated at wavevectors in the
along-flow direction.

\paragraph{Advective Form (ASC)}
\label{sec:orga2cbf9a}
On a B-type staggering the advection of the horizontal velocity by the vertical velocity requires averaging of the vertical velocity to cells.
The reconstruction operator from vertices to cells is denoted by $\mathrm{Av_v^c}$:
\begin{equation}\label{eq:vertical-advection-horizontal-momentum}
\Big( \bsw\partial_z\boldsymbol{u}\Big)_c = \Big( \mathrm{Av_v^c} w\Big)_c \partial_z \mathbf{u}_c.
\end{equation}
Due to the vanishing vertical velocity in the background flow the only contribution in the linearization is the vertical advection of the background horizontal velocities; expressed in wavenumber space: $U_z \mathsf{A_x^u}\Hat{w}$ and $V_z \mathsf{A_y^u}\Hat{w}$ where the averaging operators satisfy $\mathsf{A_x^u}=\mathsf{A_y^u}=\mathsf{Av_v^c}$.

The following simple scheme that computes the streamline directional derivative using the discrete vector gradient (short: \textbf{ASC}) might be used for the momentum advection by the
horizontal velocity field.
First the full gradient of the horizontal
velocity is computed on cells where it is dotted with the horizontal velocity to obtain the
change in velocity along a streamline:
\begin{equation*}
\label{eq:org9a01329}
\Big(\big(\boldsymbol{u}\cdot \nabla\big)\boldsymbol{u} \Big)_{c}
= \Big(\boldsymbol{u}\cdot \big(\nabla \boldsymbol{u}\big)\Big)_c.
\end{equation*}
The Fourier symbol of its linearization around a flow with constant
velocities on horizontal planes is skew-hermitian but in general not diagonal. In
particular, the addition of constant flow will yield a non-diagonal
contribution to the system matrix and, therefore, impedes Galilean invariance.
Introducing a notation for the $2\times 2$ block submatrices the Fourier symbol is given as follows:
\begin{equation}
\begin{split}
U\mathsf{G_x} + V\mathsf{G_y}&=
\renewcommand{\arraystretch}{2.0}
\left[
\begin{array}{cc}
   U\mathsf{G}_x^{(xx)} + V\mathsf{G}_y^{(xx)}  & U\mathsf{G}_x^{(xy)} + V\mathsf{G}_y^{(xy)}  \\
   U\mathsf{G}_x^{(yx)} + V\mathsf{G}_y^{(yx)}  & U\mathsf{G}_x^{(yy)} + V\mathsf{G}_y^{(yy)} 
\end{array}
\right]
\renewcommand{\arraystretch}{1.0}\\
&=
\begin{bmatrix}
& U g_x^d + V g_y^d & & \\
U g_x^u + V g_y^u & & & \\
& & & U g_x^d + V g_y^d \\
& & U g_x^u + V g_y^u &
\end{bmatrix}.
\end{split}
\end{equation}

\paragraph{Flux Form}
\label{sec:orgda3ce60}
We begin with a remark. In the continuous case, because of continuity $\nabla\mathbf{u}+\partial_z w=0$, the momentum advection $(\mathbf{u}\cdot\nabla)\mathbf{u}+w\partial_z\mathbf{u}$ can be written as $\nabla\cdot(\mathbf{u u})+\partial_z(w\mathbf{u})$. This is not longer possible on triangular B grids if the momentum advection is computed on vector control volumes as the discrete continuity is expressed for scalar control volumes. Therefore, the flux form is always taken as $\nabla\cdot(\mathbf{u u})-\mathbf{u}\nabla\cdot\mathbf{u}+\partial_z(w\mathbf{u})-\mathbf{u}\partial_zw$,\footnote{If $f(\mathbf{v}, u_i) = \nabla\cdot(\mathbf{u} u_i)+\partial_z(w u_i)$ with $i=1,2$ are components of the discrete momentum advection then the expression should be interpreted as $f(\mathbf{v}, \mathbf{u}) - \mathbf{u}f(\mathbf{v}, 1)$.} which is simply a different way of implementing the advective form. The terms with the minus sign (momentum advection due to a divergence in the velocity field on cells) sum to zero if the discrete momentum advection is first calculated on scalar control volumes and then averaged to cells.

In analogy to the advective form the flux form of the advection of horizontal velocity by the vertical velocity requires the reconstruction of the vertical velocity to cells by $\mathrm{Av_v^c}$:
\begin{equation}
    \Big( \partial_z \big(\bsw\boldsymbol{u}\big)\Big)_{c} = \frac{\partial \big(\mathrm{Av_v^c} w\big)_c \mathbf{u}_c }{\partial z}
\end{equation}
In the linearization and considering the subtraction of the momentum advection due to a divergence in the velocity field on cells the remaining contributions are the vertical advection of the background horizontal velocities as above; expressed in wavenumber space: $U_z \mathsf{Av_v^c}\Hat{w}$ and $V_z \mathsf{Av_v^c}\Hat{w}$. 

The discretization of the horizontal momentum flux through the
lateral faces of the cell can be (1) either computed on scalar
control volumes and then averaged to cells, or (2) directly computed
on cells in which case two reconstructions for the \textit{advecting}
and \textit{transported} horizontal velocities onto the faces of the
cell must be specified.
In both cases the explicit expressions for the resulting Fourier symbols, $\mathsf{G_x}$ and $\mathsf{G_y}$, are given in the accompanying notebook.

\begin{itemize}
\item \textbf{Divergence on Scalar Control Volumes (FDV).}
\label{sec:org4a66bd0}
The divergence of momentum flux is first 
computed on each scalar control volume. It is then averaged back to the
cells using the same averaging operator $\mathrm{Av_v^c}$ as above:
\begin{align*}
 \Big(\nabla \cdot \big( \boldsymbol{u} u_i \big) \Big)_{c}
 = \Big(\mathrm{Av_v^c}\big(\nabla\cdot \big(\boldsymbol{u} u_i)\big) \Big)_c && i=1,2.
\end{align*}
In this case the induced advection due to a divergence in the velocity field on cells vanishes due to continuity on median-dual control volumes.
Analyzing the Fourier symbol of the linearized operator, it
should be noted that the off-diagonal submatrices vanish \(\mathsf{G}_a^{(xy)} =
\mathsf{G}_a^{(yx)} = \boldsymbol{0}\) for both \(a=x,y\) but the diagonal
submatrices, $\mathsf{G}_a^{(xx)}$ and $\mathsf{G}_a^{(yy)}$ with $a=x,y$, are not diagonal in general but skew-hermitian.

\item \textbf{Divergence on Cells with Linear Vector Reconstructions onto Edges (FDCRE).}
\label{sec:org97ef672}
The horizontal velocity is linearly reconstructed at each triangle. At each edge of the cell the volume flux through the respective edge is
computed and multiplied with an upwind velocity reconstruction giving the momentum flux advected through the edge:
\begin{subequations}
\label{eq:org2f9c0cd}
\begin{align*}
F_{e, \Tilde{c}} &= \ell_e \sum_{\hat{c}\in C(e)} \boldsymbol{n}_{\Tilde{c},e}\cdot\boldsymbol{u}_{\hat{c}} / 2\\
R_{e, \Tilde{c}}u_i &= u_{i,\Tilde{c}} + d_e/2~\boldsymbol{n}_{\Tilde{c},e} \cdot\big(\nabla u_i\big)_{\Tilde{c}}\\
\Big(\nabla \cdot \big( \boldsymbol{u} u_i \big) \Big)_{c}
&= \frac{1}{A_c}\sum_{e\in C(e)} \sum_{\Tilde{c}\in C(e)} \boldsymbol{n}_{c,e}\cdot\boldsymbol{n}_{\Tilde{c},e}\frac{F_{e,\Tilde{c}} +  \abs{F_{e,\Tilde{c}}}}{2}~R_{e,\Tilde{c}}u_i
\end{align*}
\end{subequations}
In this case too the off-diagonal submatrices in the Fourier symbol of
the linearized operator vanish but the diagonal submatrices are
full. The physical branch of the eigenvalues is more accurate than for
the other schemes. There is, however, a spurious/geometric mode in
the opposite direction of the advecting background flow (see Figure \ref{fig:hmt-FS}).
\end{itemize}

\paragraph{Vector-Invariant Form (AVI)}
\label{sec:org5f92ac0}
Another approach is to discretize the vector-invariant formulation of
the momentum transport (short: \textbf{AVI}). The tangential component of the momentum flux
may be discretized by computing the curl on each median-dual control volume
and reconstruct the results to the cells where they are multiplied by the horizontal velocity rotated by $90^\circ$. The component of the momentum flux
in flow direction may be approximated by calculating the gradient of
the kinetic energy on each cell. The kinetic energy on the median-dual
control volumes may be obtained by the reconstruction operator, $\mathrm{Av_c^v}$, applied to the horizontal velocity on cells:
\begin{multline*}
\label{eq:org2a0b994}
\Big(\big(\boldsymbol{u}\cdot \nabla\big)u_i \Big)_{c}
= \bigg( \mathrm{Av_v^c}\Big( \mathrm{Curl}~\boldsymbol{u} \Big) \left(\boldsymbol{u}^\perp\right)_i \bigg)_c\\ + \frac{1}{2}\boldsymbol{e}_i\cdot\bigg(\nabla\left( \mathrm{Av_c^v}\big(\boldsymbol{u}\big) \cdot \mathrm{Av_c^v}\big(\boldsymbol{u}\big)\right)  \bigg)_c.
\end{multline*}
In the Fourier symbol of the linearized operator the submatrices
\(\mathsf{G}_a^{(xx)}\) and \(\mathsf{G}_a^{(yy)}\) are full for both \(a=x,y\). In addition
the cross-interaction submatrices \(\mathsf{G}_x^{(yx)}\) and \(\mathsf{G}_y^{(xy)}\) are
non-zero and real which indicates the possibility of spurious instabilities along the symmetric axis (\textit{generalized Hollingsworth instabilities}).  Interestingly, the following relations between the Fourier
symbols in the different schemes hold.
\begin{align*}
\mathsf{G}_x^{(xx)}(\mathrm{AVI}) &= \mathsf{G}_x^{(xx)}(\mathrm{FDV}) + \mathsf{G}_y^{(xy)}(\mathrm{AVI})\\ \mathsf{G}_x^{(yy)}(\mathrm{AVI}) &= \mathsf{G}_x^{(yy)}(\mathrm{FDV})\\
\mathsf{G}_y^{(yy)}(\mathrm{AVI}) &= \mathsf{G}_y^{(yy)}(\mathrm{FDV}) + \mathsf{G}_x^{(yx)}(\mathrm{AVI})\\ \mathsf{G}_y^{(xx)}(\mathrm{AVI}) &= \mathsf{G}_y^{(xx)}(\mathrm{FDV})
\end{align*}
The explicit expressions for the Fourier symbols $\mathsf{G_x}(\mathrm{AVI})$ and $\mathsf{G_y}(\mathrm{AVI})$ are given in the accompanying notebook.

\subsubsection{Transport of Buoyancy}
Since the vertical velocity and buoyancy fields are collocated at the vertices on a triangular grid with B-type staggering, no reconstructions in the horizontal direction are needed to compute the vertical advection of buoyancy in this case.
The horizontal buoyancy transport is expressed in terms 
of buoyancy fluxes across the
bounding faces of the median-dual control volume, \(v\), as follows:
\begin{equation}
A_v\Big(\nabla \cdot \big(\boldsymbol{u}\bsb\big) \Big)_{v}
=\sum_{e\in E(v)} d_e \boldsymbol{n}_{v,e} \boldsymbol{u}_e b_e.
\end{equation}
The advecting horizontal velocity $\mathbf{u}_e$ is taken in the same form as in the continuity equation (mean over triangles on both sides of the respective edge). 
The second order method is obtained by the central reconstruction of buoyancy to mid-edges \(b_e = \sum_{v\in V(e)} b_v/2\). 
A fourth-order method can be obtained in several ways. The simplest way is to apply $(1-(1/8)a^2\Delta)$ to the buoyancy field before the reconstruction (averaging) to mid-edge (see, e.g. \cite{Smolentseva2020}). 
Here, $a$ is the length of triangle side and $\Delta$ is the discrete scalar Laplacian operator defined in Equation \eqref{eq:TriB-laplacian}. Other ways will introduce similar corrections but with directional Laplacians and would imply more work. 

The Fourier symbol of the linearized advection operator acting on the
buoyancy may be expressed in terms of three contributions of different
origin.
\begin{itemize}
\item \textbf{The vertical advection of mean buoyancy}:
Since the vertical velocity and buoyancy are collocated at the
vertices of the triangular grid, no additional averaging other than in
vertical direction due to the Lorenz-type staggering is needed. The
contribution to the Fourier symbol involves the Brunt-Väisälä
frequency of background flow and acts on the amplitude, $\Hat{w}$, that specifies the vertical velocity field as a plane wave: \(N^2\Hat{w}\).
\item \textbf{The horizontal advection of mean buoyancy}:
Due to the staggering of the horizontal velocity with respect to
buoyancy, the horizontal velocity requires averaging in order to
compute the directional derivative of mean buoyancy along the velocity
field. In fact, on a triangular grid it turns out that there are
cross-interactions; i.e. the $x$-component of the horizontal velocity
also advects the change of mean buoyancy in the $y$-direction and vice
versa. In particular, in the Eady configuration the background
velocity contributes to the advection of mean buoyancy too. 
The explicit expressions for the averaging operators' Fourier symbols, $\mathsf{A_x^b}$ and $\mathsf{A_y^b}$, acting on the amplitudes of the horizontal velocity fields are provided in the accompanying notebook.
As a result the contribution to the Fourier symbol acts on the four amplitudes that specify the horizontal velocity field as a plane wave and has the following form: $\big(B_x \mathsf{A_x^b} + B_y \mathsf{A_y^b}\big) \mathbf{\Hat{u}}$.

\item \textbf{The horizontal advection of buoyancy by the background flow}:
\label{sec:orgc8db8bb}
The use of an higher-order accurate reconstruction scheme for the
buoyancy yields an improved accuracy for the buoyancy's rate of change
along the flow direction. Figure \ref{fig:hstb-FS} illustrates the difference
in the Fourier symbols evaluated at wavevectors in the flow direction
of the respective schemes. However, due to the staggering of
horizontal velocity and buoyancy the Fourier symbol differs from the
Fourier symbol in the advection of horizontal momentum (\(Ug_x + Vg_y\))
and will instead be expressed in terms of a new symbol for the gradient denoted by $\mathsf{\Gamma}$ (see accompanying notebook for the explicit expression).
The resulting contribution to the Fourier symbol of the linearized advection operator acting on the amplitude, $\Hat{b}$, specifying the buoyancy field as a plane wave is then given by the following expression: \(\left( U\mathsf{\Gamma}_x + V \mathsf{\Gamma}_y \right) \Hat{b}\).
\end{itemize}

\begin{figure*}
    \begin{subfigure}{0.5\linewidth}
    \centering
    \includegraphics[width=0.99\linewidth]{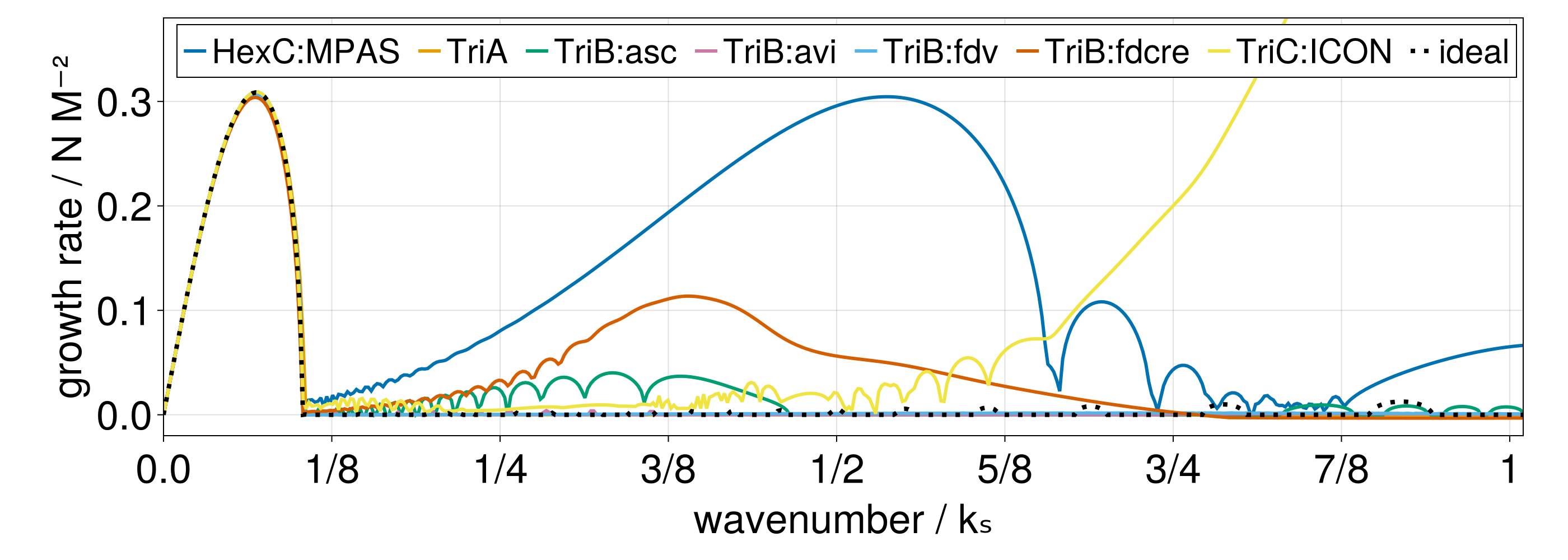}
    \end{subfigure}
    ~
    \begin{subfigure}{0.5\linewidth}
    \centering
    \includegraphics[width=0.99\linewidth]{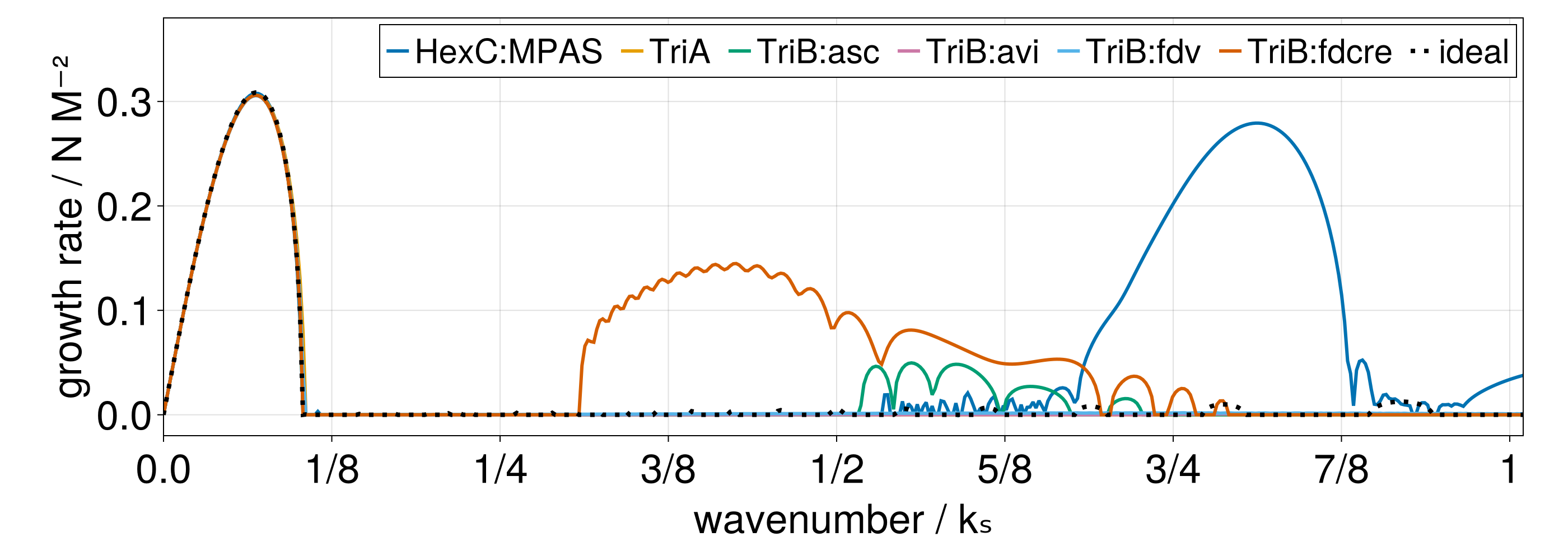}    
    \end{subfigure}
    \begin{subfigure}{0.5\linewidth}
    \centering
    \includegraphics[width=0.99\linewidth]{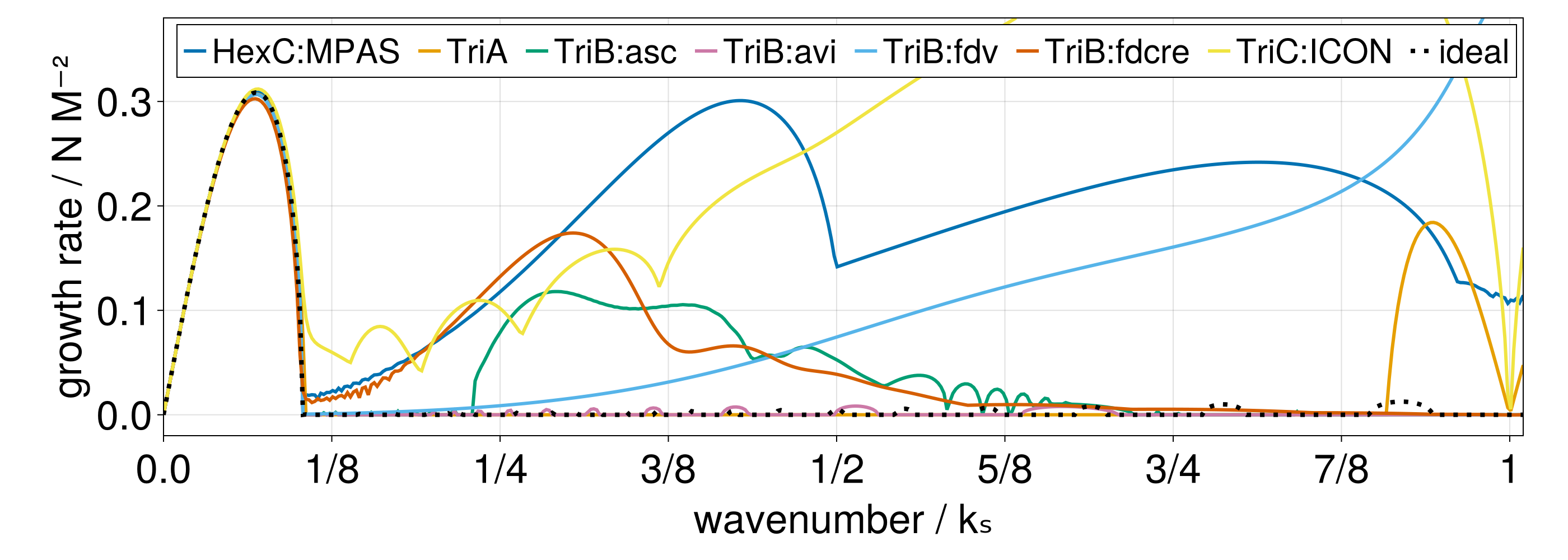}
    \end{subfigure}
    ~
    \begin{subfigure}{0.5\linewidth}
    \centering
    \includegraphics[width=0.99\linewidth]{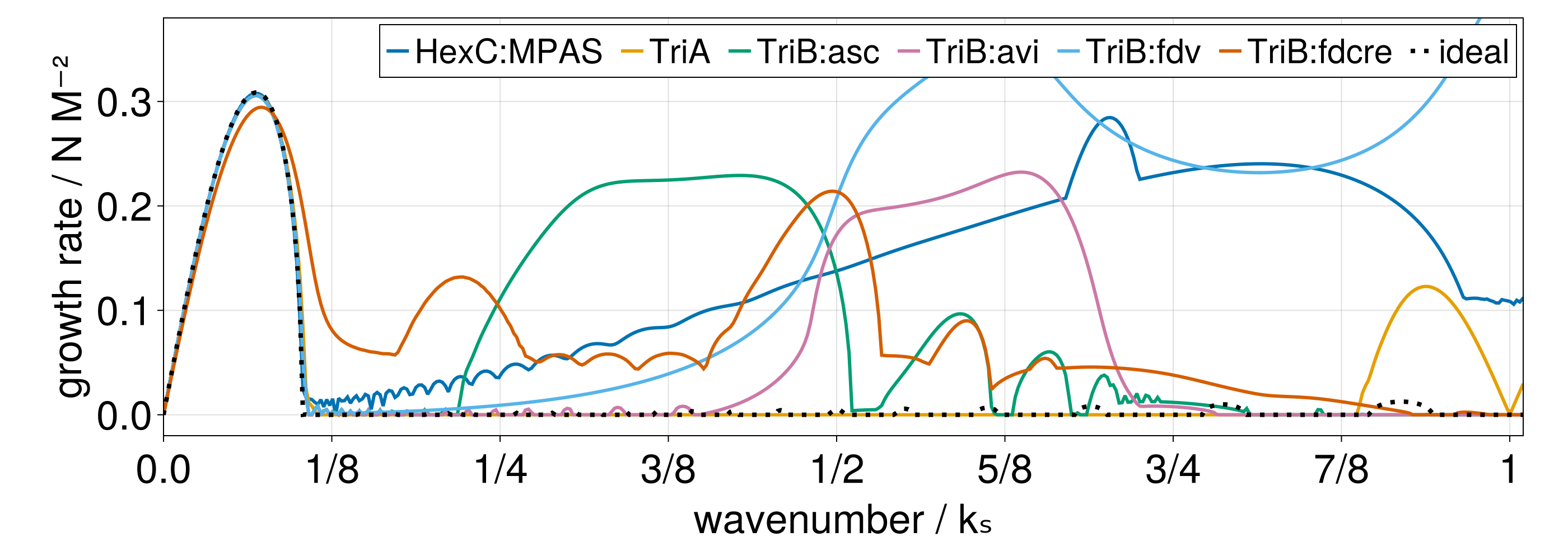}    
    \end{subfigure}
    \begin{subfigure}{0.5\linewidth}
    \centering
    \includegraphics[width=0.99\linewidth]{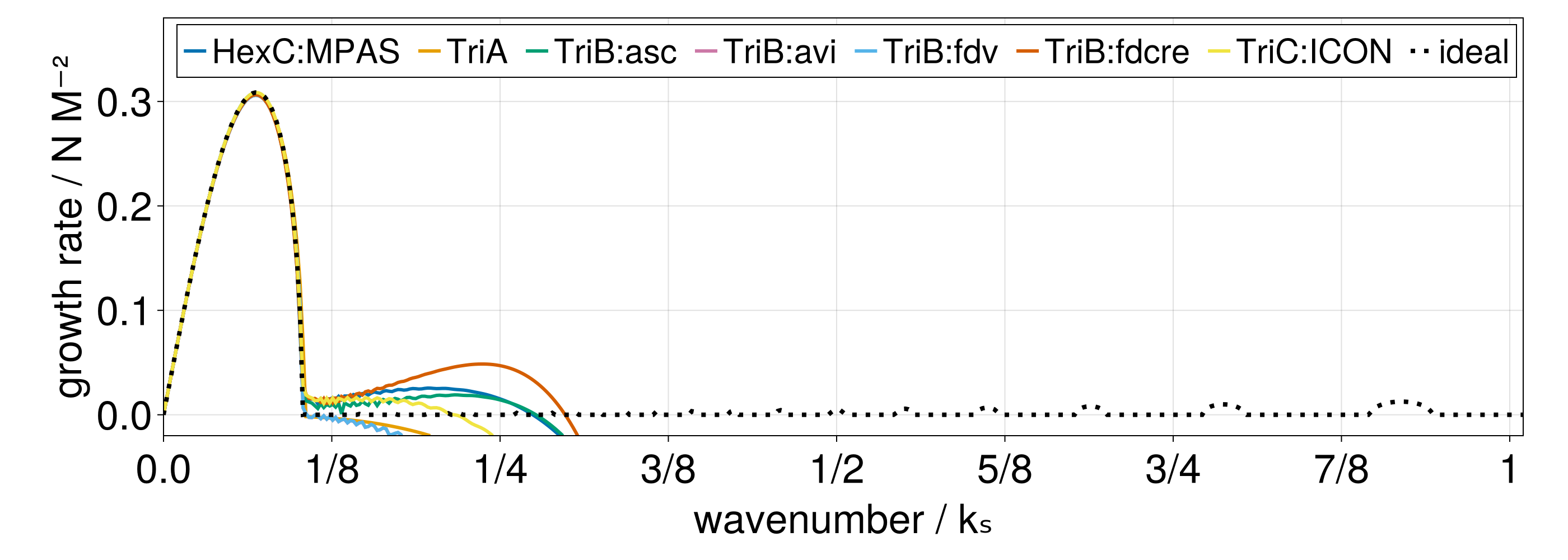}
    \end{subfigure}
    ~
    \begin{subfigure}{0.5\linewidth}
    \centering
    \includegraphics[width=0.99\linewidth]{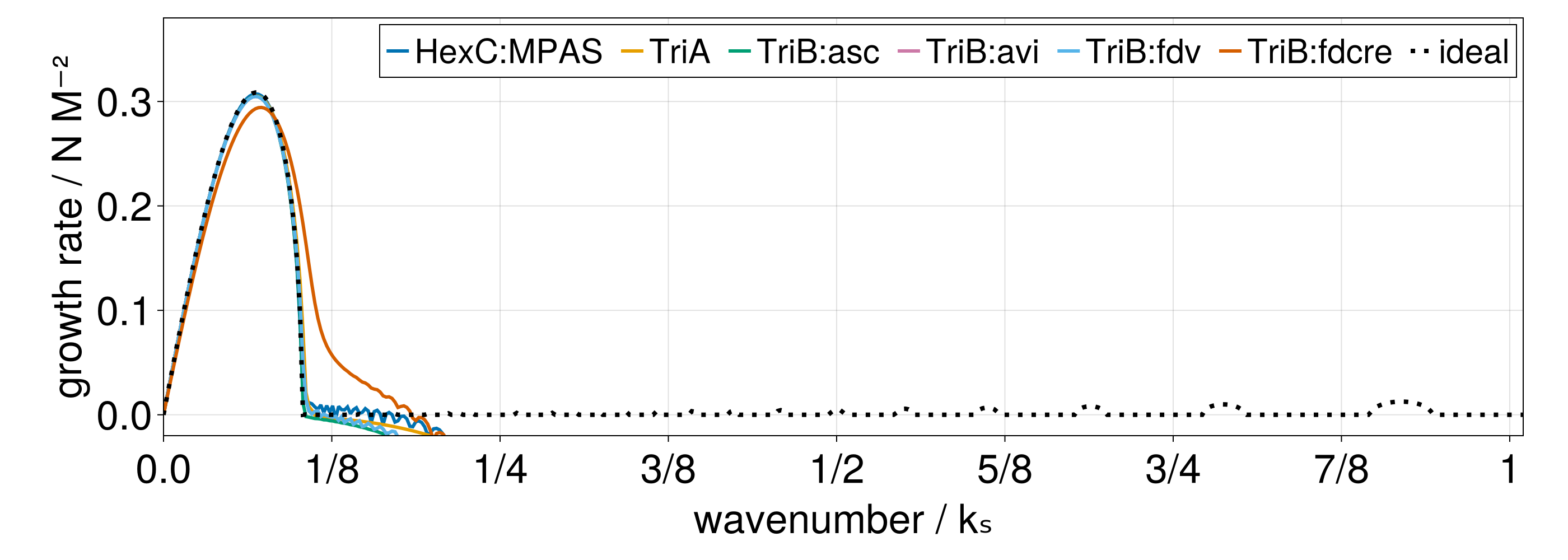}    
    \end{subfigure}
    \begin{subfigure}{0.5\linewidth}
    \centering
    \includegraphics[width=0.99\linewidth]{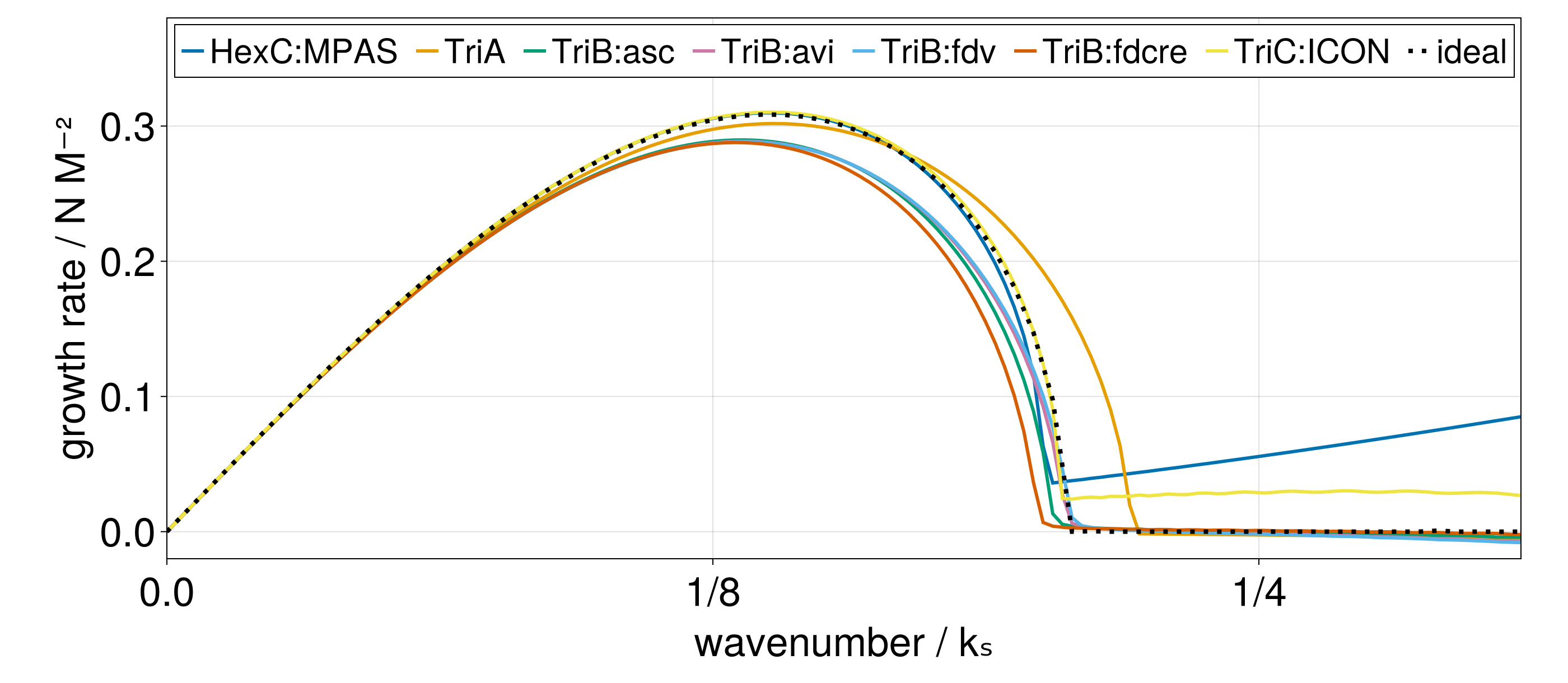}
    \end{subfigure}
    ~
    \begin{subfigure}{0.5\linewidth}
    \centering
    \includegraphics[width=0.99\linewidth]{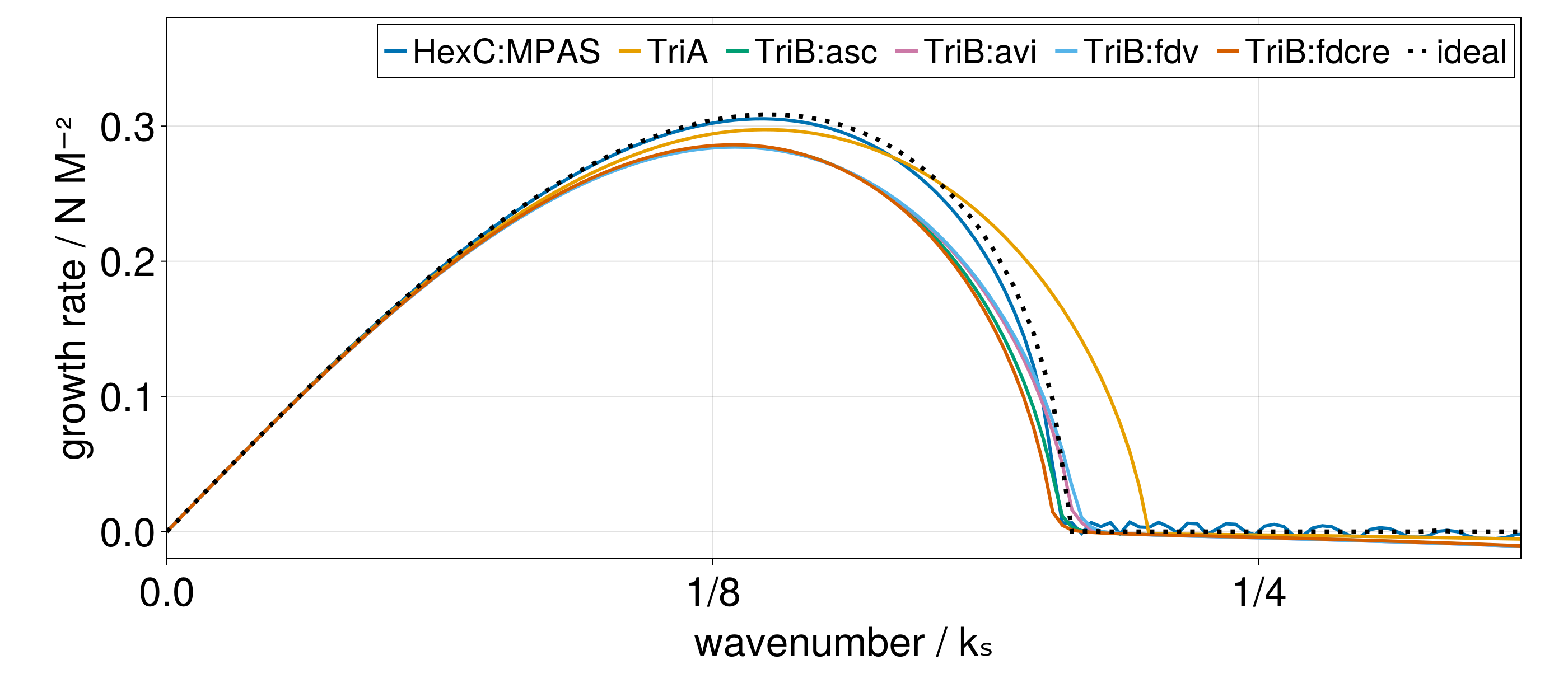}    
    \end{subfigure}
    \caption{Growth rate of the fastest growing instability on a triangular and hexagonal grids as a function of wavenumber on the baroclinic axis. The Richardson number is $\mathrm{Ri}=100$, the Brunt-Väisälä frequency $N=10^{-3} ~s^{-1}$, the Coriolis parameter $f_0=-10^{-4}~s^{-1}$, the number of vertical layers $N_z=64$, the total depth $H=4000~m$, the lattice parameter $a=6250~m$ (first three rows) and $a=12500~m$ (fourth row).
    \textit{Left column}: Low-order accurate advection schemes. \textit{Right column}: High-order accurate advection schemes (not implemented for C grid). \textit{First row}: Azimuth $\theta_U=0 ~\mathrm{rad}$, $\beta=0$, no active dissipation. \textit{Second row}: Azimuth $\theta_U=\pi/6~\mathrm{rad}$, $\beta=0.5$, no active dissipation. \textit{Third row}: Azimuth $\theta_U=\pi/6~\mathrm{rad}$, $\beta=0.5$, biharmonic friction $\mathbb{V}=0.01~ms^{-1}$. \textit{Fourth row}: Azimuth $\theta_U=0 ~\mathrm{rad}$, $\beta=0$, biharmonic friction with $\mathbb{V}=0.001~ms^{-1}$.}
    \label{fig:bc}
\end{figure*}

\begin{figure*}
    \begin{subfigure}{0.99\linewidth}
    \centering
    \includegraphics[width=0.7\linewidth]{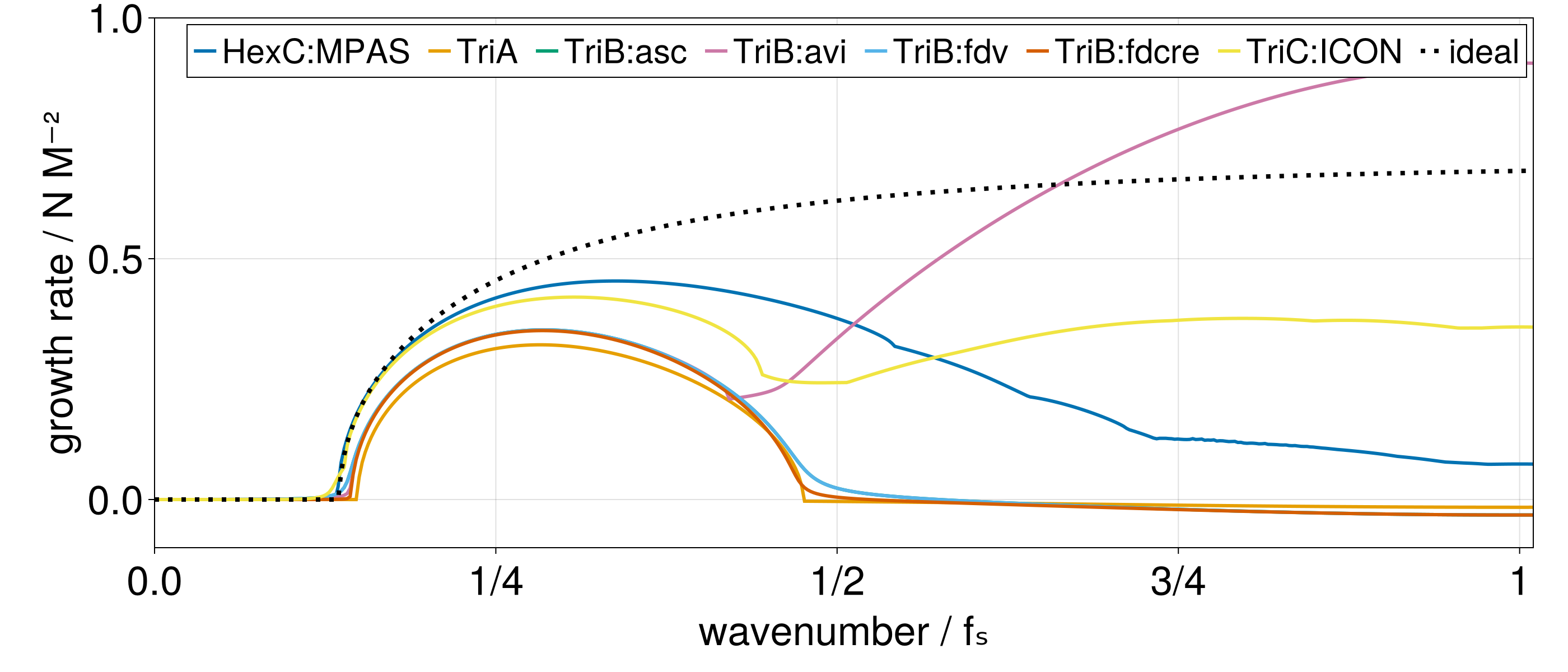}
    \end{subfigure}
    \begin{subfigure}{0.99\linewidth}
    \centering
    \includegraphics[width=0.7\linewidth]{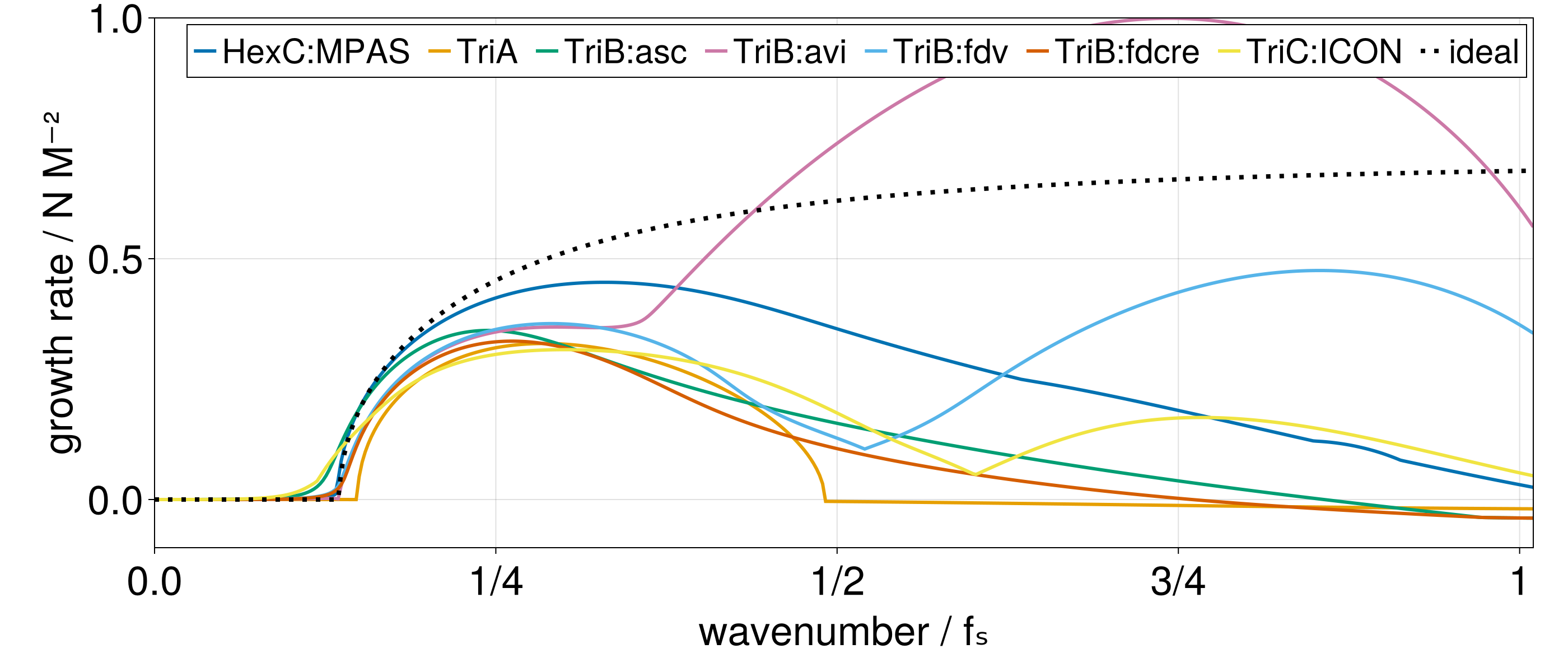}
    \end{subfigure}
    \begin{subfigure}{0.99\linewidth}
    \centering
    \includegraphics[width=0.7\linewidth]{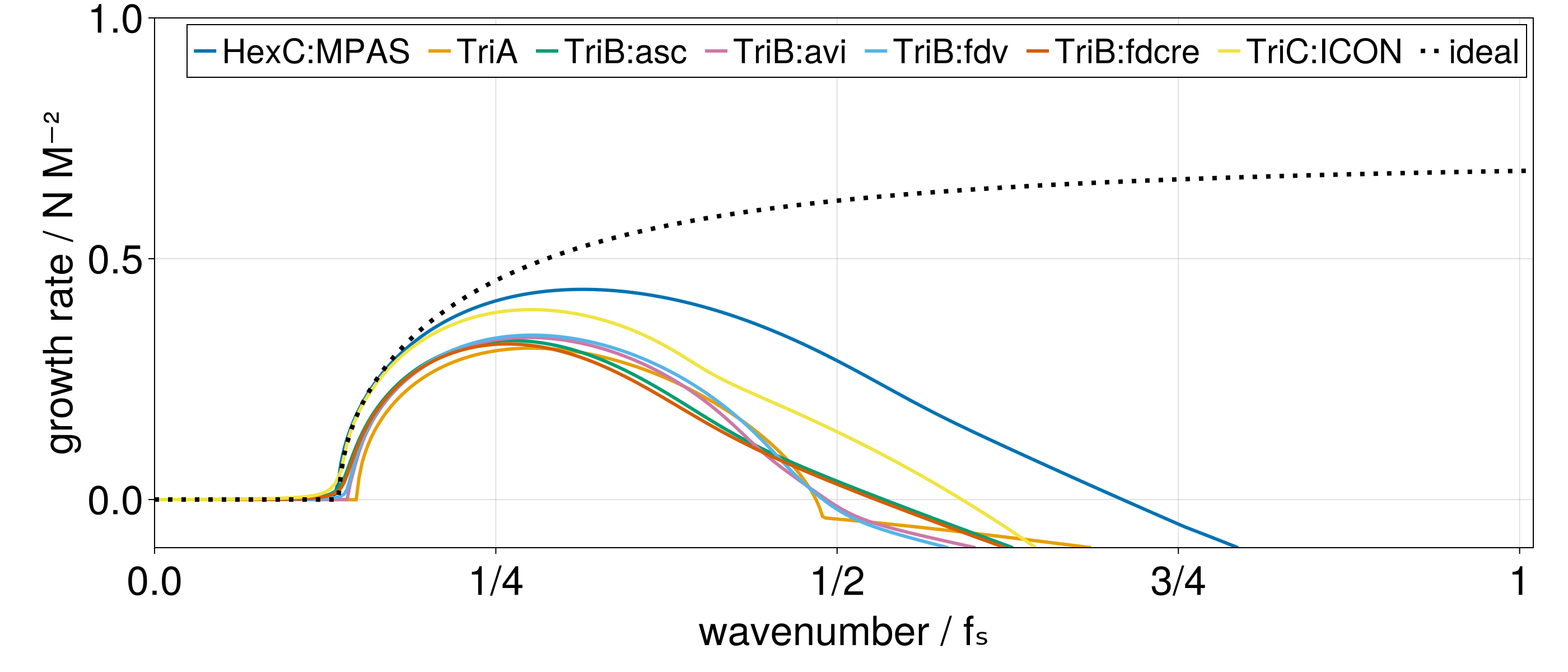}
    \end{subfigure}
    \begin{subfigure}{0.99\linewidth}
    \centering
    \includegraphics[width=0.7\linewidth]{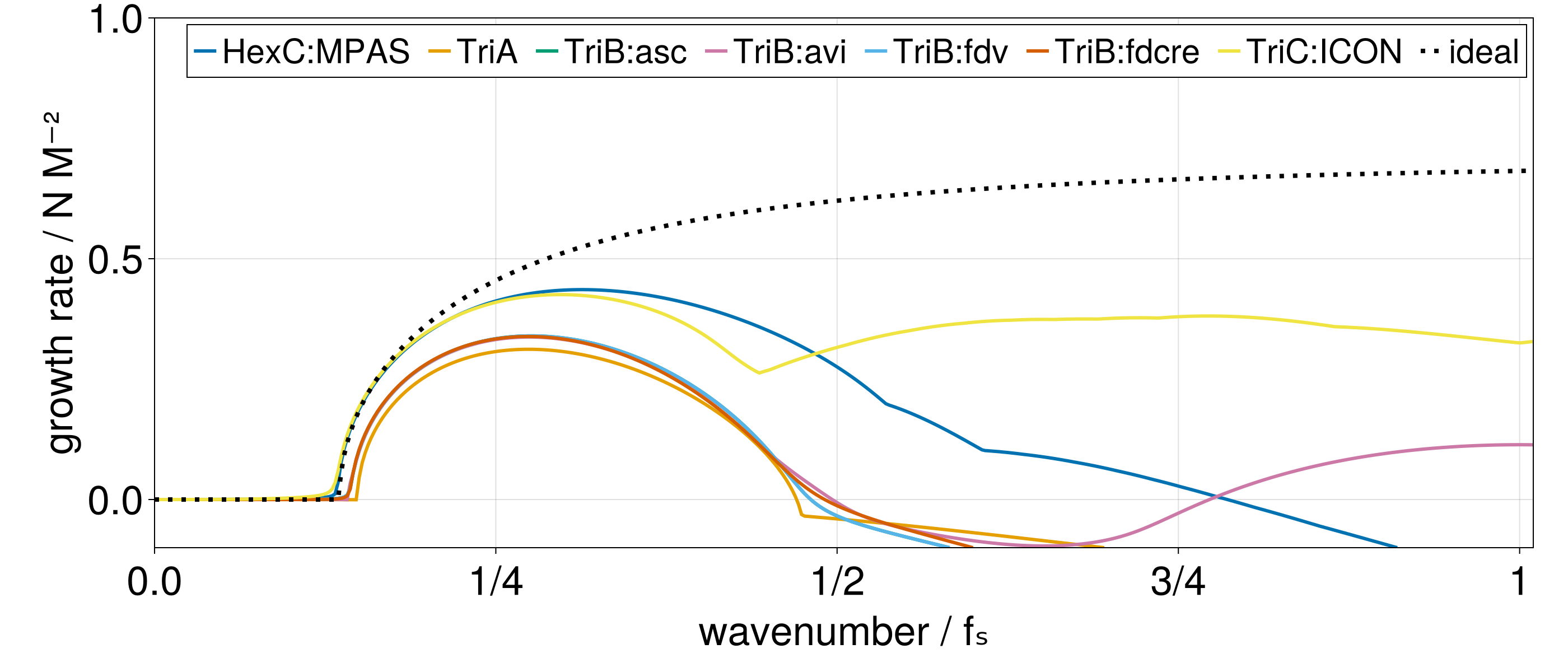}
    \end{subfigure}
    \caption{Growth rate of the fastest growing instability on a triangular and hexagonal grids as a function of wavenumber on the symmetric axis. The Richardson number is $\mathrm{Ri}=0.5$, the Brunt-Väisälä frequency $N=10^{-3} ~s^{-1}$, the Coriolis parameter $f_0=-10^{-4}~s^{-1}$, the number of vertical layers $N_z=64$, the total depth $H=4000~m$, the lattice parameter $a=6250~m$. \textit{First row}: Azimuth $\theta_U=0 ~\mathrm{rad}$, $\beta=0$, biharmonic friction with $\mathbb{V}=0.001~ms^{-1}$. \textit{Second row}: Azimuth $\theta_U=\pi/6~\mathrm{rad}$, $\beta=0.5$, , biharmonic friction with $\mathbb{V}=0.001~ms^{-1}$. \textit{Third row}: Azimuth $\theta_U=\pi/6~\mathrm{rad}$, $\beta=0.5$, biharmonic friction with $\mathbb{V}=0.01~ms^{-1}$. \textit{Fourth row}: Azimuth $\theta_U=0~\mathrm{rad}$, $\beta=0.5$, biharmonic friction with$\mathbb{V}=0.01~ms^{-1}$.}
    \label{fig:sy}
\end{figure*}

\subsection{Triangular Grid with C-type Staggering}\label{sec:method-C-type}
In a triangular C grid formulation the scalar quantities are located
at the circumcenters of the respective triangles. The finite volume
interpretation of their values is the mean on the respective triangle
(cell). The dual mesh is a Voronoi diagram where each vertex of the
triangulation is the collocation point of a Voronoi cell (also called
\textit{dual cell}). Only diagnostic quantities are evaluated at the
grid points of the dual mesh in the primitive model; in particular,
the vorticity is computed on the Voronoi cells. The grid points
of the horizontal velocity components are staggered with respect to the
triangle circumcenters and with respect to each other. They are
located at the midpoints of the (primal) grid edges. The value of the
horizontal velocity component on each such point is interpreted as the
mean of the normal component of the horizontal velocity on the
respective edge (with a given orientation). See the right panel of Figure \ref{fig:staggering} for an illustration of the C-type staggering on a triangular mesh.

A horizontal discretization on a triangular grid with C-type
staggering has been formulated by Korn in \cite{Korn2017a} and is
implemented in the ocean component of
the ICON model. The discretization approach is based
on a weak formulation of the equations of motion and the
discretization schemes of the differential operators apply the
\textit{mimetic approach}.
The weak formulation of the momentum balance uses a non-canonical choice of scalar product on the discrete space of normal velocities on edges that yields a mass matrix. In practice, a simple lumping method is applied to the mass matrix.

\subsubsection{Discrete Linear Operators}
Using once again the notation as summarized in Table \ref{tab:notation-Tri} the discrete analogs of the gradient of a scalar field, the divergence of a flux field, and the curl of the horizontal velocity field are given by the following expressions:

\begin{subequations}
    \begin{equation}
        \Big( \nabla \bsp \Big)_e = \frac{1}{d_{e}} \sum_{c\in C(e)} -\boldsymbol{n}_{c,e}^T\boldsymbol{n}_e p_c\label{eq:TriC-gradient}
    \end{equation}

    \begin{equation}
        \Big( \nabla \cdot \boldsymbol{F} \Big)_c = \frac{1}{A_c}\sum_{e\in E(c)} \ell_e \boldsymbol{n}_{e,c}\cdot\boldsymbol{n}_e F_e\label{eq:TriC-divergence}
    \end{equation}
    \begin{equation}
        \Big( \mathrm{Curl}~\boldsymbol{u} \Big)_v = \frac{1}{A_v} \sum_{e\in E(v)} -\ell_e \boldsymbol{n}_{v,e} \cdot \boldsymbol{t}_e u_e\label{eq:TriC-curl}
    \end{equation}
    \begin{equation}
        \Big( \Delta\bsb \Big)_c = \Big(\nabla\cdot \nabla\bsb\Big)_c\label{eq:TriC-laplacian}
    \end{equation}
    \begin{equation}
        \Big( \Delta\bsu \Big)_e = \Big(\nabla \big(\nabla\cdot\bsu\big)\Big)_e + \frac{1}{\ell_e}\sum_{v\in V(e)} \mathbf{n}_{e,v}^T \mathbf{t}_e \big(\mathrm{Curl}~\bsu\big)_v.\label{eq:TriC-vector-laplacian}
    \end{equation}
\end{subequations}

The discrete gradient operator acting on the pressure field simply evaluates the centered finite difference across each edge in the direction defined by the orientation of the edge, $\boldsymbol{n}_e$.
Here the edges corresponding to the same lattice are given the same orientation:
\begin{align*}
    \boldsymbol{n}_{e^{(1)}} = \begin{bmatrix}\sqrt{3}/2\\1/2\end{bmatrix} && \boldsymbol{n}_{e^{(2)}} = \begin{bmatrix}-\sqrt{3}/2\\1/2\end{bmatrix} &&
    \boldsymbol{n}_{e^{(3)}} = \begin{bmatrix}0\\1\end{bmatrix}.
\end{align*}

The Fourier symbol of the discrete gradient operator acts on the two amplitudes that specify the pressure field as a plane wave with fixed wavevector, $\boldsymbol{k}$. Thus, the matrix, $\mathsf{G}$, has size $3\times 2$ and satisfies the following relation
\begin{equation}
\begin{bmatrix}
    e^{-i\boldsymbol{k}\cdot\boldsymbol{x}_{e^{(1)}}} \Big( \nabla \boldsymbol{p} \Big)_{e^{(1)}}\\
    e^{-i\boldsymbol{k}\cdot\boldsymbol{x}_{e^{(2)}}} \Big( \nabla \boldsymbol{p} \Big)_{e^{(2)}}\\
    e^{-i\boldsymbol{k}\cdot\boldsymbol{x}_{e^{(3)}}} \Big( \nabla \boldsymbol{p} \Big)_{e^{(3)}}
\end{bmatrix}
    = 
    \frac{1}{d_e}
    \begin{bmatrix}
        -e^{-i\alpha_1/2} & e^{i\alpha_1/2}\\[0.6em]
        -e^{-i\alpha_2/2} & e^{i\alpha_2/2}\\[0.6em]
        -e^{-i\alpha_3/2} & e^{i\alpha_3/2}
    \end{bmatrix}
    \begin{bmatrix}\Hat{p}_{c^u}\\\Hat{p}_{c^d}\end{bmatrix}
    =\mathsf{G} \begin{bmatrix}\Hat{p}_{c^u}\\\Hat{p}_{c^d}\end{bmatrix}.
\end{equation}

The gradients of kinetic energy, pressure, and surface elevation in the momentum balance are discretized by an application of the discrete gradient operator followed by a filtering operator that results from the special scalar product in the weak formulation.
This filtering operator is, however, eliminated in the lumping process of the mass matrix.

The Fourier symbol of the discrete divergence operator may be expressed in terms of the gradient's Fourier symbol: $\mathsf{D} = -4 \mathsf{G}^*/3$ where $\mathsf{G}^*$ denotes the Hermitian transpose of $\mathsf{G}$.
In the discretization of the continuity equation \eqref{eq:TriC-continuity-eq} the discrete divergence operator is preceded by the application of a filtering operator introduced next:
\begin{equation}
    \left(\nabla\cdot \left(\Pi^T\Pi \boldsymbol{u}\right) \right)_{c} + \partial_z w_{c} = 0.\label{eq:TriC-continuity-eq}
\end{equation}
The inclusion of the filtering operator is a natural consequence of the weak formulation. It may be interpreted as a requirement to satisfy energy consistency (cf. \cite{Korn2017a}).

The discrete curl operator acting on the horizontal velocity field uses Stokes' theorem to approximate the curl on the dual cells \eqref{eq:TriC-curl}.
The Fourier symbol of the curl has size $1\times 3$ and is denoted by $\mathsf{C}$.
The discrete Laplace operator on cells \eqref{eq:TriC-laplacian} is the divergence of the scalar field's gradient. 
Thus, its Fourier symbol is the product of the Fourier symbols of the divergence and gradient operators:
\begin{equation}
    \mathsf{L^b} = \mathsf{DG} = -\frac{4}{3} \mathsf{G^* G}.
\end{equation}
In particular, the Fourier symbol is Hermitian negative definite but in general not diagonal and same holds for the symbol of biharmonic diffusion, $\mathsf{D^b}=-\mathbb{K}^b\mathsf{L}^b\mathsf{L}^b$, for positive diffusion coefficients, $\mathbb{K}^b>0$.

The discrete vector Laplacian on edges \eqref{eq:TriC-vector-laplacian} has two terms. 
The first term is the gradient of the horizontal velocity's divergence defined on cells.
Its Fourier symbol is the product of the Fourier symbols of the gradient and divergence operators: $\mathsf{GD} = -4~\mathsf{D^*D}/3$.
The second term is the horizontal gradient field of the vorticity rotated by $90^\circ$.
Its Fourier symbol, denoted by $\mathsf{L^u}$, can also be shown to be Hermitian negative definite and same carries over for the symbol of biharmonic viscosity, $\mathsf{D^u}=-\mathbb{K}^u\mathsf{L}^u\mathsf{L}^u$, for positive viscosity coefficients, $\mathbb{K}^u>0$.
The explicit expression of the Fourier symbols are given in the accompanying notebook.

In order to discretize the advection operators where products of scalar and vector fields need to be computed the C grid formulation of ICON-o \cite{Korn2017a} requires an operator that reconstructs vector fields to cells.
Its transpose reconstructs vector fields on cells to edges and the composition of the two operators must be a positive-definite filtering operator on the space of vector fields (on edges).
The Perot operator is used, as introduced in \cite{Perot2000}:
\begin{align}
    \Big( \Pi \boldsymbol{u} \Big)_c
&= \frac{1}{A_c} \sum_{e\in E(c)} \frac{\ell_e d_e}{2} \boldsymbol{n}_{c,e}\cdot\boldsymbol{n}_e \boldsymbol{n}_{c,e} u_e\\
\Big( \Pi^T\boldsymbol{F} \Big)_e
&= \frac{1}{2} \sum_{c\in C(e)} \boldsymbol{n}_e \cdot \boldsymbol{F}_c.
\end{align}
Moreover, the computation of the Coriolis force in the momentum balance requires an operator that reconstructs vector fields to vertices and another operator that reconstructs the rotation of a vector field defined on vertices by $90^\circ$ to edges.
The resulting Coriolis operator is energy-neutral if the composition of the two operators is skew-symmetric on the space of vector fields (on edges).
To distinguish these reconstruction operators from the Perot operators the notations $\Tilde{\Pi}$ and $\Tilde{\Pi}^\dagger$ are used:
\begin{align}
    \Big( \Tilde{\Pi} \boldsymbol{u} \Big)_v 
    &= \frac{1}{A_c} \sum_{e\in E(v)} \frac{\ell_e d_e}{2} \boldsymbol{n}_{e,v}^\bot\cdot \boldsymbol{n}_e \boldsymbol{n}_{e,v}^\bot u_e\\
    \Big( \Tilde{\Pi}^\dagger \boldsymbol{G}\Big)_e
&= \frac{1}{2} \sum_{v\in V(e)} -\boldsymbol{t}_e \cdot \boldsymbol{G}_v
\end{align}
The interested reader can find explicit expressions of the Fourier symbols, $\mathsf{\Pi}$, $\mathsf{\Tilde{\Pi}}$, and $\mathsf{\Tilde{\Pi}^\dagger}$, corresponding to the respective reconstruction operators in the accompanying notebook.

\subsubsection{Transport of Horizontal Momentum}
In the advective formulation of the horizontal momentum transport in the vertical direction on a triangular grid with C-type staggering the discretization requires averaging due to the staggering of the horizontal velocities with respect to the vertical velocity.
The product of vertical velocity with the derivative of the horizontal velocity in the vertical direction is computed on cells using a Perot reconstruction of the horizontal velocities.
The product is then reconstructed to the edges by the transpose Perot reconstruction:
\begin{equation}
    \big( \bsw\partial_z \boldsymbol{u}\big)_e = \Big(\Pi^T \big(w \partial_z\big(\Pi \mathbf{u}\big)\big)\Big)_e
\end{equation}
In the linearization the only contribution is given by the vertical advection of the background horizontal velocities; expressed in wavenumber space: $\left(U_z  \mathsf{A_x^b}+ V_z \mathsf{A_y^b}\right) \Hat{w}$ with $\mathsf{A_x^b} = \mathsf{\Pi_x^{T}}$ and $\mathsf{A_y^b}=\mathsf{\Pi_y^{T}}$.

\begin{figure*}
    \begin{subfigure}{0.32\linewidth}
        \centering
        \includegraphics[width=\linewidth]{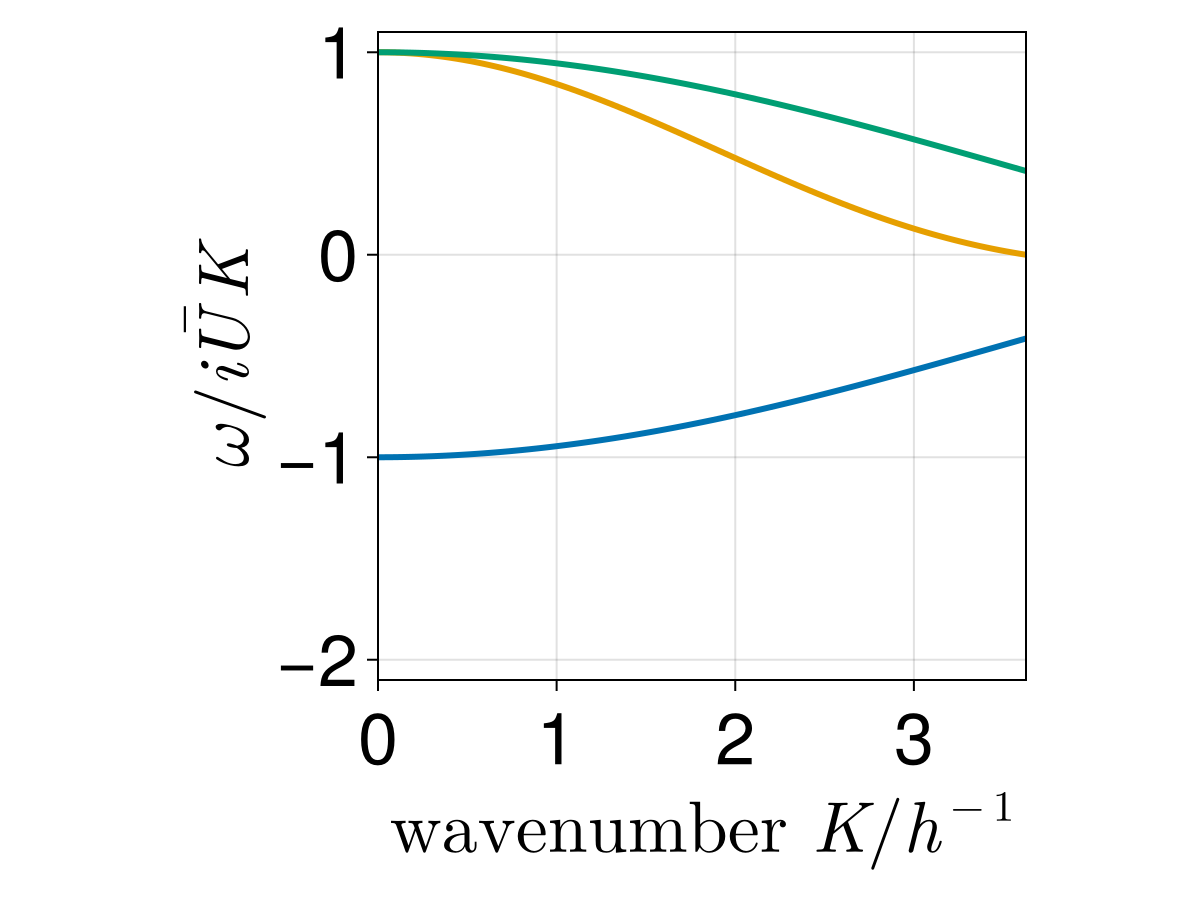}
        \phantomcaption
         \label{fig:TriC-hmt-FS}
    \end{subfigure}
    \begin{subfigure}{0.32\linewidth}
        \centering
        \includegraphics[width=\linewidth]{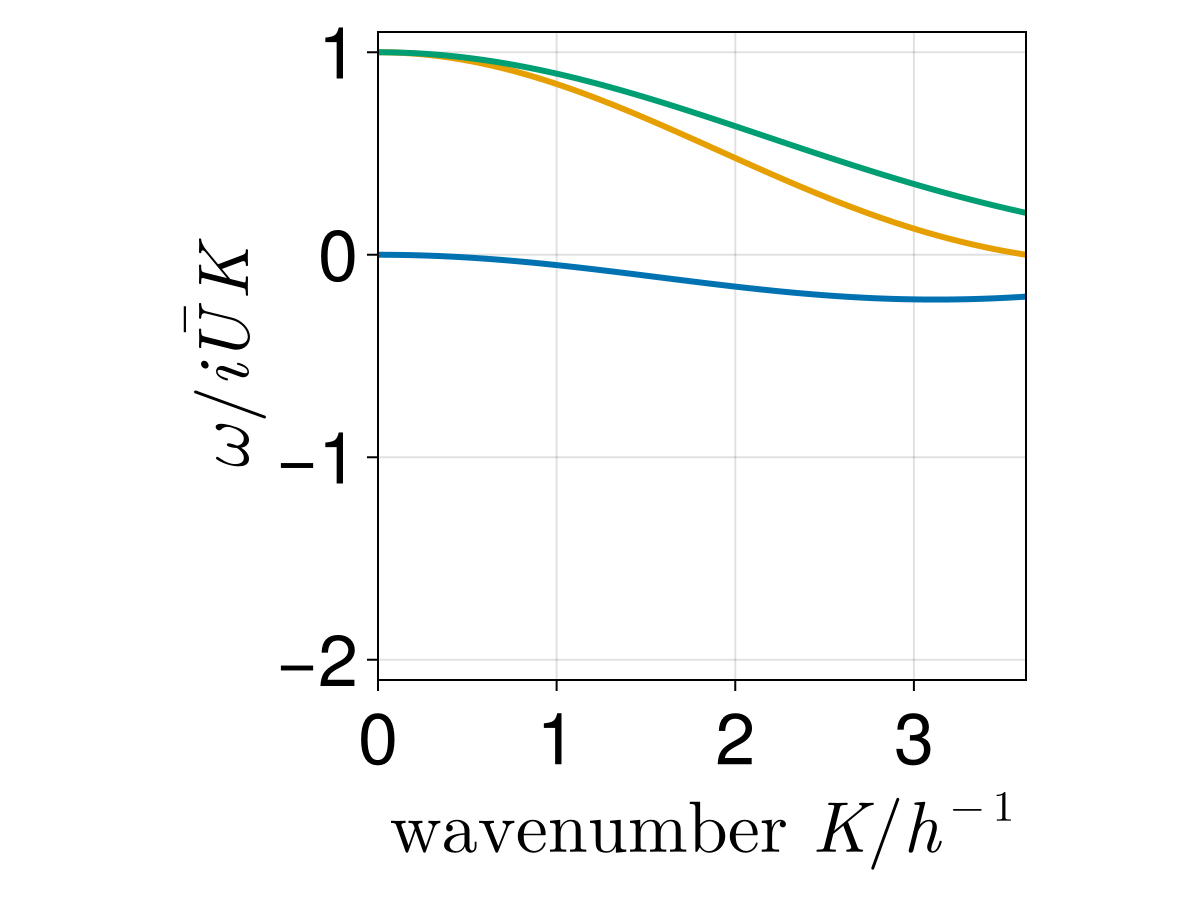}
        \phantomcaption
         \label{fig:TriC-hmt-FS-alt}
    \end{subfigure}
    \begin{subfigure}{0.32\linewidth}
        \centering
        \includegraphics[width=\linewidth]{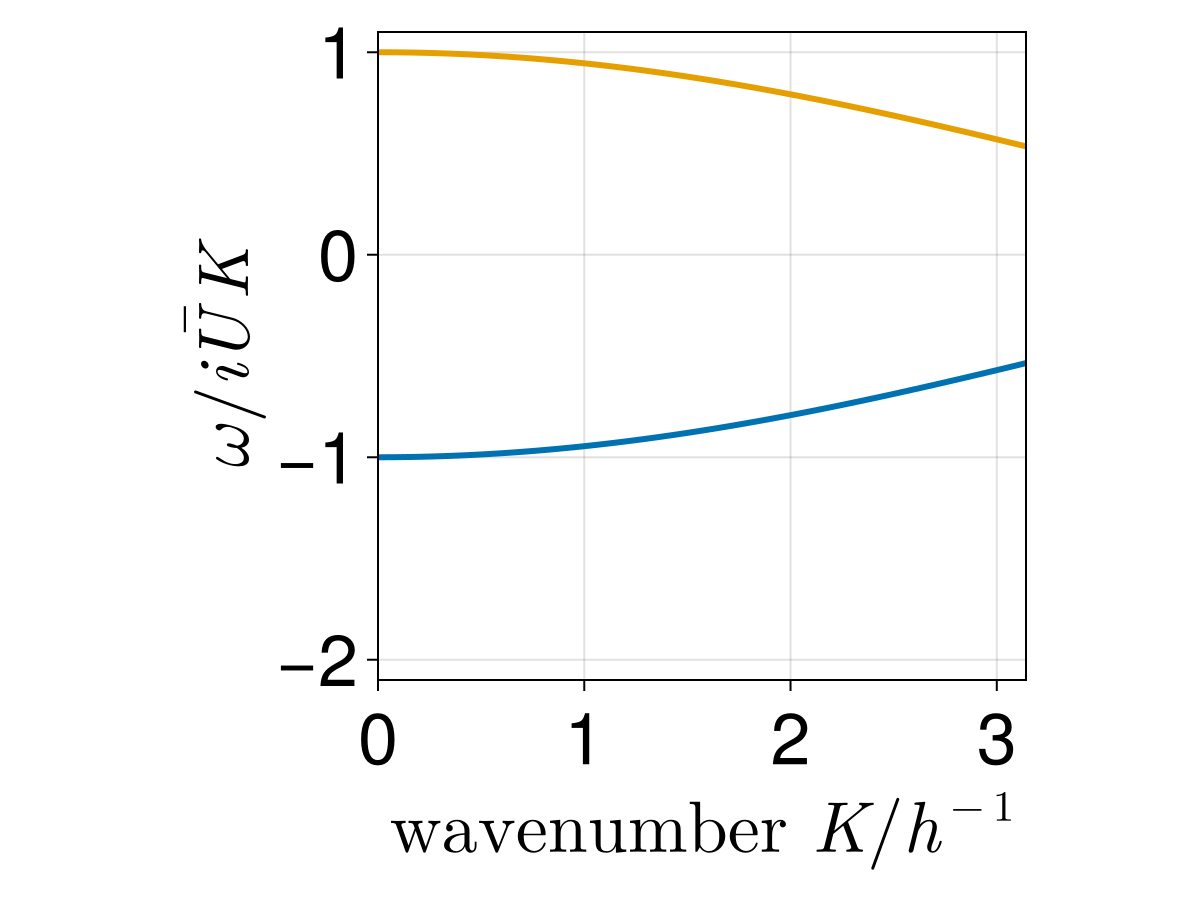}
        \phantomcaption
        \label{fig:TriC-hstb-FS}
    \end{subfigure}
    \caption{\textit{Left}: Eigenvalues of the linearized discrete momentum advection operator's Fourier symbol on triangular C grid as function of the wavenumber in the flow direction. There are two  physical branches with different accuracy coming from two parts of vector invariant momentum advection. The overall accuracy is similar to the accuracy of the ASC scheme for B-type staggering. Further, the symbol maintains a spurious branch with opposite direction of propagation. \textit{Middle}: Same as \textit{left} but including the filtering operator acting on the gradient of kinetic energy. \textit{Right}: Fourier Symbol of the linearized advection schemes acting on buoyancy as a function of the wavenumber in the flow direction. }\label{fig:TriC-adv-ops}
\end{figure*}

The scheme for the horizontal momentum transport in the horizontal direction is
based on the vector-invariant formulation. 
The kinetic energy is computed on cells after an application of the Perot reconstruction to the horizontal velocity field.
As was mentioned before the filtering operator, $\Pi^T\Pi$, that would follow the gradient operator acting on kinetic energy is canceled due to the mass matrix.
The flux of the the velocity field rotated by its curl
is approximated by multiplying the curl on dual cells with the Perot
reconstructed velocity field and then apply the Perot operator
$\Tilde{\Pi}^\dagger$:
\begin{subequations}
\begin{align*}
\mathrm{KE}_c &= \left(\Pi\boldsymbol{u}\right)_c \cdot \left(\Pi\boldsymbol{u}\right)_c\\
\omega_v &= \mathrm{Curl}(\boldsymbol{u})\\
\Big(\big(\boldsymbol{u}\cdot \nabla\big)\boldsymbol{u} \Big)_e
&= \left(\Tilde{\Pi}^\dagger \left(\boldsymbol{\omega} \Tilde{\Pi}\boldsymbol{u}\right) \right)_e + \left( \nabla \boldsymbol{\mathrm{KE}} \right)_e
\end{align*}
\end{subequations}
The Fourier symbol of the discrete advection operator linearized around a flow with constant velocity on horizontal planes acts on the three amplitudes that specify the horizontal velocity field as a plane wave with fixed wavevector. 
The Fourier symbol satisfies the following relation:
\begin{equation*}\label{eq:TriC-adv-mom}
    U \mathsf{G_x} + V\mathsf{G_y}
    = U \left( \mathsf{\Tilde{\Pi}^{\dagger}_x}\mathsf{C} + \mathsf{G}\mathsf{\Pi_x}  \right) + V \left( \mathsf{\Tilde{\Pi}_y^{\dagger}}\mathsf{C} + \mathsf{G}\mathsf{\Pi_y} \right).
\end{equation*}
Its eigenvalues when evaluated at wavevectors in the direction of the flow are shown in Figure \ref{fig:TriC-hmt-FS}.
The explicit expression of the Fourier symbol can be further analyzed in the accompanying notebook.

\subsubsection{Transport of Buoyancy}
Since the vertical velocity and buoyancy fields are collocated at the cells on a triangular grid with C-type staggering, no reconstructions in the horizontal direction are needed to compute the vertical advection of buoyancy.
The horizontal velocity field is reconstructed on cells by the Perot operator to compute the advective flux of buoyancy in the horizontal direction on cells.
In order the compute the divergence of the flux on cells by the operator in Equation \eqref{eq:TriC-divergence}, the advective flux is then reconstructed to edges by the transpose Perot operator.
\begin{equation}
\left(\nabla \cdot \big(\boldsymbol{u}\bsb\big) \right)_c
= \left(\nabla\cdot \left(\Pi^T \left( \bsb\Pi\boldsymbol{u} \right)\right) \right)_c
\end{equation}
The Fourier symbol of the linearized discrete advection operator is split into three contributions:
\begin{itemize}
\item \textbf{The vertical advection of mean buoyancy}:
Since the vertical velocity and buoyancy are collocated at the
cells of the triangular grid, no additional averaging other than in
vertical direction due to the Lorenz-type staggering is needed. The
contribution to the Fourier symbol involves the Brunt-Väisälä
frequency of background flow and acts on the amplitudes, $\Hat{w}$, that specify the vertical velocity field as a plane wave: $N^2\Hat{w}$.
\item \textbf{The horizontal advection of mean buoyancy}:
Due to the staggering of the horizontal velocity with respect to the
buoyancy the horizontal velocity requires averaging in order to
compute the directional derivative of mean buoyancy along the velocity
field. 
This part of the Fourier symbol acts on the three amplitudes that specify the horizontal velocity field as a plane wave and has the following form: $(B_x \mathsf{A_x^b} + B_y \mathsf{A_y^b}) \mathbf{\Hat{u}}$.
The explicit expressions for the average operators' Fourier symbols may be found in the accompanying notebook.
\item \textbf{The horizontal advection of buoyancy by the background flow}:
Due to the staggering of
horizontal velocity and buoyancy the Fourier symbol differs from the
Fourier symbol in the advection of horizontal momentum (\(Ug_x + Vg_y\))
and will instead be expressed in terms of a new symbol for the gradient denoted by $\mathsf{\Gamma}$ (see accompanying notebook for the explicit expression).
It turns out that these new derivative symbols $\mathsf{\Gamma_x}$ and $\mathsf{\Gamma_y}$ coincide with the components of the gradient's Fourier symbol defined on cells for a B-type staggering \eqref{eq:TriB-gradient-vector}.
The resulting contribution to the Fourier symbol of the linearized advection operator acting on the amplitude, $\Hat{b}$, specifying the buoyancy field as a plane wave is given by: $\left(U \mathsf{\Gamma_x} + V\mathsf{\Gamma_y}\right) \Hat{b}$.
\end{itemize}

\subsection{Implementation and System Matrix}

\begin{table*}
\begin{minipage}{0.5\linewidth}
    \centering
    \begin{tabular}{lrrr}
    \toprule
    Staggering & A & B & C\\[0pt]
    \midrule
    quadrilateral mesh & 2 & 2 & 2\\[0pt]
    triangular mesh & 2 & 4 & 3\\[0pt]
    \bottomrule
    \end{tabular}
\end{minipage}
\begin{minipage}{0.49\linewidth}
    \centering
    \begin{tabular}{lrrr}
    \toprule
    Staggering & A & B & C\\[0pt]
    \midrule
    quadrilateral mesh & 1 & 1 & 1\\[0pt]
    triangular mesh & 1 & 1 & 2\\[0pt]
    \bottomrule
    \end{tabular}
\end{minipage}
\caption{Number of amplitudes to determine the horizontal velocity field (\textit{left}) and a scalar quantity field (\textit{right}) in wavenumber space of a horizontal plane.}
\label{tab:amplitudes}
\end{table*}

Following the approach from Barham et al. \cite{Barham2018} the vertical velocity and pressure variables are eliminated by the introduction of the operators $\mathsf{\underline{W}}$ and $\mathsf{\underline{P}}$ respectively. This is essentially achieved by integrating the discrete version of the continuity equation \eqref{eq:continuity-eq} (resp. hydrostatic balance equation \eqref{eq:hydrostatic-eq}) in wavenumber space and applying the boundary condition of vanishing vertical velocity at the bottom surface \eqref{eq:bottom-bc} (resp. vanishing hydrostatic pressure at the top \eqref{eq:top-bc}). Expressions for the corresponding matrices can be found in the appendix \eqref{eq:Wop} and \eqref{eq:Pop}.
Note that these operators are defined on the tensor product of the respective horizontal space with the vertical space.
Operators between two tensor products of a discrete horizontal with the vertical space will be denoted by an underlining.
For notational convenience the underlining of an operator between two discrete horizontal spaces denotes the tensor product of the operator with the identity on the vertical space (i.e. Kronecker product of the matrices).

The fully assembled system matrices for the considered discretization on a triangular grid with B-type staggering and C-type staggering are summarized in Equation \eqref{eq:TriB-sys-mat-alt}. All parameters of the system are summarized in Table \ref{tab:parameters}.
The system matrix acts on the vector of amplitudes specifying the horizontal velocity fields on the different vertical layers, the buoyancy field on the different vertical layers, and the surface elevation.

\begin{figure*}[h!]
\renewcommand{\arraystretch}{1.4}
\begin{equation}
\mathsf{S}=
\begin{bmatrix}
\mathsf{G_x}\otimes\underline{U} + \mathsf{G_y}\otimes\underline{V} + \left( \mathsf{A_x^u}\otimes\underline{U_z} + \mathsf{A_y^u}\otimes\underline{V_z}\right) \mathsf{\underline{W}} + f_0 \mathsf{\underline{M}} + \mathsf{\underline{D^u}} & \mathsf{\underline{G}}~\mathsf{\underline{P}} & g  \mathsf{G\otimes 1_V}\\
N^2 \mathsf{\underline{W}} + \mathsf{A_x^b}\otimes\underline{B_x} + \mathsf{A_y^b}\otimes\underline{B_y}& \mathsf{\Gamma_x}\otimes\underline{U} + \mathsf{\Gamma_y}\otimes\underline{V} +  \mathsf{\underline{D^b}}& \underline{0}\\
\Delta_z \mathsf{D} \otimes \mathsf{1_V^T} & \mathsf{0 \otimes 1_V^T} & \mathsf{0}
\end{bmatrix}
\label{eq:TriB-sys-mat-alt}
\end{equation}
\renewcommand{\arraystretch}{1.0}
\end{figure*}
In the case of the triangular B grid discretization with linear upwind reconstruction of the transported horizontal velocity to edges (FDCRE) the expression of the system matrix in Equation \eqref{eq:TriB-sys-mat-alt} is formally correct if the background flow is unidirectional.
If the velocity profile is symmetric around mid-level then the Fourier symbols, $\mathsf{G_x}$ and $\mathsf{G_y}$, in the horizontal momentum advection depend on the vertical component so that the linearized horizontal momentum advection operator is not a simple tensor in this case.

In the Supplementary Material for each Fourier symbol an expression that may be interpreted as function in the \textit{Julia} programming language is provided together with a notebook for assembling the system matrix.
The notebook further allows for symbolic analysis and manipulations of the Fourier symbols for the different variable staggerings including A-type staggering on a triangular mesh and C-type staggering on a hexagonal mesh not discussed in the main text.
Replacing a symbol by its exact value (i.e. the limit for $a\to0$), the influence of the corresponding operator and its accuracy may analyzed using our setup.

\section{Discussion}
Spatial discretizations used in atmospheric and ocean modeling are
commonly discussed in the context of shallow water equations (see
e.g.\cite{Randall1994,LeRoux2007,Yu2020,Lapolli2024}). In real-world
applications, additional factors such as advection of scalars and
momentum as well as vertical discretization come into play. In the
vertical direction, ocean circulation models traditionally use the so
called Lorenz staggering (see discussion in \cite{Bell2017b})
independent of the horizontal discretization. This vertical staggering
is known to lead to spurious instability of computational kind
(\cite{Bell2017b}, \cite{Menesguen2025}) if the vertical
resolution is too fine compared to the horizontal one.
The horizontal discretization of momentum advection may trigger the
Hollingsworth instability for certain variants of the vector-invariant
form (\cite{Hollingsworth1983, Bell2017a, Ducousso2017}). This
illustrates that the accuracy and stability of particular spatial
discretizations result from intricate interaction of many factors,
and their analysis needs a framework that is less limiting than
linearized shallow water equations.

\paragraph{Analysis Framework: Discrete Eady Problem}
The ability of particular discretizations to approximate the
theoretical growth rate of baroclinic and symmetric instability in the
Eady configuration depends on all factors mentioned above. As shown by
\cite{Barham2018}, the respective analysis presents a framework for
learning how particular features of discretizations influence their
performance. For instance, \cite{Barham2018} show that on
quadrilateral meshes the C grid is more accurate than the B grid in
the representation of symmetric instability, and is only slightly more
accurate in the representation of the growth rate of baroclinic
instability.  An unexpected finding of \cite{Barham2018} is a spurious
instability along the baroclinic axis on quadrilateral C grids. While
\cite{Barham2018} show that on sufficiently fine meshes this
instability is easily suppressed by moderate biharmonic viscosity and
diffusion if a fourth-order advection of buoyancy is used, much higher
dissipation is needed on coarse meshes because the spurious instability
penetrates into the range of wavenumbers of physical baroclinic
instability.

\paragraph{Extension of the Analysis to Triangular and Hexagonal Grids}
Several new ocean circulation models (FESOM, ICON-o and MPAS-Ocean)
are formulated on unstructured (triangular or hexagonal) meshes and
use different horizontal spatial discretizations. An obvious
motivation for studying the discrete Eady problem for these
discretizations is the question of their accuracy. Some aspects of the
answer can be foreseen, as C grids on such meshes generally imply less
averaging than B grids, same as on quadrilateral meshes. 
The averaging operators acting on the vertical velocity in the horizontal momentum equation as well as the averaging operators acting on the horizontal velocity in the buoyancy advection stipulate the accuracy of the instabilities' growth rate especially on the symmetric axis. 
Reflected in the accuracy of the Fourier symbols of these operators, $\mathsf{A}$ and $\mathsf{Av}$, the discretization schemes on triangular grids with B-type staggering require more averaging than schemes with C-type staggering.
More important, however, is the question about the consequences of spurious
numerical modes (see e.g. \cite{LeRoux2007,LeRoux2012,Danilov2013})
maintained by staggered discretizations on triangular or hexagonal
meshes. These modes were studied predominantly in the context of
linearized shallow water equations, where they create spurious wave
branches which are neutral and hence can be damped. A real question is
what happens in unstable 3-dimensional flows, especially in the light of possible
spurious baroclinic instability found in \cite{Barham2018}. 
Our study reveals many unexpected details which are briefly discussed
below but a complete analysis of all branches of spurious instabilities shown in Figure \ref{fig:bc} on the baroclinic and in Figure \ref{fig:sy} on the symmetric axis are beyond the scope of this work. 
We finish with several remarks on the quadrilateral meshes which we further explored for comparison.

\subsection{Instabilities on Triangular and Hexagonal Meshes}
Formulated briefly, the main result of our analysis in this work is
that all staggered discretizations (i.e. B and C grids) maintain
spurious instabilities along baroclinic and symmetric axes. All need
dissipation to suppress these modes. The spurious part is generally a
combination of several branches, and we hypothesize that they are
partly related to the spurious modes maintained by these
discretizations. The reader shall be reminded that these modes are
created by the disagreement between the numbers of discrete velocity
and pressure degrees of freedom, which is due to the geometry of
triangular or hexagonal meshes. 
On quadrilateral meshes the velocity/pressure ratio of degrees of freedom in the horizontally
discretized problem is approximately 2 to 1. The use of triangular and hexagonal meshes,
however, entails a mismatch in these degrees of freedom (see Figure \ref{fig:staggering} and Table
\ref{tab:amplitudes}). Therefore, the geometry of these meshes allows for spurious numerical modes. In fact, the advection operators on triangular grids
with B-type staggering (see Figure \ref{fig:TriB-adv-ops}) and C-type staggering (see Figure \ref{fig:TriC-adv-ops}) do maintain spurious numerical modes. 

Comparing the first and second row in Figure \ref{fig:bc} (Figure \ref{fig:sy}) the spurious branches of baroclinic (resp. symmetric) instabilities show a strong dependence on the azimuth of the background flow direction in the coordinate system of the grid as well as the speed of the background flow at mid-level.
A higher-order reconstruction of the buoyancy field in the horizontal advection yields more accurate symbols, $\mathsf{\Gamma_x}$ and $\mathsf{\Gamma_y}$, for the gradient in the linearization.
As a consequence, on the baroclinic axis the spurious branches of instabilities show a significant improvement (see the right column of Figure \ref{fig:bc}).
In particular, the physical and spurious branches of instabilities show some separation in wavenumber on the baroclinic axis.

\paragraph{Alignment of the Flow with the Mesh and Violation of Galilean Invariance}
In contrast to the quadrilateral meshes, the symmetric axis also
maintains spurious instability branches. There is strong sensitivity of the
spurious branches to the direction of the flow with respect to the mesh.

Additionally, spurious instabilities violate Galilean invariance,
being sensitive to the addition of a uniform flow to the background flow centered
around the mid-depth. These sensitivities are illustrated in the second row in Figure \ref{fig:bc} and \ref{fig:sy} on the baroclinic and symmetric axis respectively. 
This is an indication that the spurious
instabilities involve spurious modes. A possibility for this violation
is created by the representation of discrete velocities. On triangular
and hexagonal C grids, the three normal velocity projections
over-define a vector field (see,
e.g. \cite{Danilov2010,Gassmann2011}). In
this case the Fourier symbol of the linearized momentum advection has
two physical and one spurious eigenvalue (Figure \ref{fig:TriC-hmt-FS} and Figure \ref{fig:TriC-hmt-FS-alt}). For triangular B grids,
there are two spurious eigenvalues due to the geometrical difference
in the orientation of any two adjacent triangles (Figure \ref{fig:hmt-FS}). These spurious
eigenvalues do not describe physical advection. Uncovering how the
spurious modes of these staggered discretizations induce spurious
branches of baroclinic and symmetric instabilities might be worth of
attention, but is beyond the scope of the present work.

\paragraph{Damping of Spurious Instabilities on the Baroclinic Axis}
In each case a relatively moderate dissipation is generally sufficient
to suppress the instability on the baroclinic axis. The dissipation
depends on the resolution, on the orientation of the flow to the mesh,
and on the accuracy and implementation of the advection operators.
All these factors must be taken into account when choosing appropriate viscosity and diffusivity parameters to suppress spurious branches of instabilities.
As a general rule, however, biharmonic diffusion and viscosity with coefficients scaling as
$\mathbb{V} a^3$ are generally sufficient to suppress spurious modes for
relatively moderate values of $\mathbb{V}$ ($\mathbb{V} <0.01$ m/s, except for very
coarse meshes).
But even these values may already affect the accuracy
of maximum instability increment or the range of wavenumbers where
physical instability is observed (see third and fourth row in Figure \ref{fig:bc}).
On a coarser mesh with grid parameter $a\approx 25$ km the separation of physical and spurious branches of instabilities in wavenumber requires a higher-order advection scheme in the buoyancy field, and higher viscosity and diffusivity parameter values are needed to suppress the spurious instabilities.

\paragraph{Damping of Spurious Instabilities on the Symmetric Axis}
On the symmetric axis, the spurious part is often attached to the
physical part, and it is difficult to see where the physical branch
ends and the spurious one starts. The associated growth rates depend
on the details of advection and flow direction. The analysis of
eigenvectors might be helpful to disentangle spurious and physical
parts, but it is delayed to future studies. The actual dissipation
needed to suppress spurious symmetric instabilities depends on
this analysis, but in any case, the situation is less promising than
the one on the quadrilateral meshes (see third row in Figure \ref{fig:sy}).

\subsection{Instabilities on Quadrilateral Meshes}
First, the analysis of \cite{Barham2018} assumes that the basic flow
is aligned with the mesh. The other limiting case is the flow directed
along the mesh diagonal. We found that in this case the accuracy is even
slightly improved, so that the results of \cite{Barham2018} hold for
arbitrary orientation of the flow and the mesh.

Second, the discrete linearized equations written down in
\cite{Barham2018} for B and C grids show that on the symmetric axis
the B-grid equations differ mainly by additional averaging of vertical
velocity in the vertical advection of mean zonal momentum and horizontal velocity in
the advection of mean buoyancy in the buoyancy equation. The pressure
gradient averaging on B grid does not affect perturbations along the
baroclinic or symmetric axes. This alone implies that the traditional
thinking about the B and C grids in terms of the accuracy of pressure
gradient is insufficient, and other factors are responsible for the
less accurate behavior of the B grids. To put this result in a broader
context, we analyzed the quadrilateral A grid. In this case the
averaging of vertical velocity in momentum equation drops out, but the
pressure gradient is less accurate on the baroclinic and symmetric
axes. The result (not shown here, but reproducible with the scripts in
the Supplementary Material) is that the A grid is similar to the B
grid as concerns the representation of the growth rates of baroclinic
and symmetric instabilities. The averaging leading to the loss of
accuracy (as compared to C grid) is different on A and B grids, but
leads to similar consequences. This also means that the pressure
gradient averaging which is the main drawback of A grids in the
context of shallow water wave dynamics does not really make A grids
worse than B grids. However, similar to the C grid, the A grid
maintains spurious instability on the baroclinic axis. The spurious
part is separated by a gap from the physical one at $a=12.5$ km or
higher, but merges with the physical one for $a=25$ km. Respectively,
biharmonic friction with $\mathbb{V}=0.01$ m/s is needed on the coarse mesh
to suppress the spurious branch, but already $\mathbb{V}=0.003$ m/s is
sufficient for $a=12.5$ km. The very similar behavior is found for the
C grid, where $\mathbb{V}=0.01$ m/s is needed for $a=25$ km and $\mathbb{V}=0.006$
m/s is needed for $a=12.5$ km in the case of the fourth order method
for buoyancy advection. The fourth order advection of buoyancy is
helpful on the A grid, but its presence is less critical than for C
grids.

Third, although the C grid is more accurate in representing symmetric
instability, it is not accurate enough at high wavenumbers. Since the
simulated instability range continues to the cutoff wavenumber,
additional dissipation will be needed to suppress the instability on
grid scales, as all discrete operators are inaccurate in this case.

\section{Conclusion}
As a follow-up to \cite{Barham2018} we analyzed the discrete Eady
instability problem on triangular A and B grids, for the mimetic
discretization of ICON-o (abbreviated as triangular C grid) and on the
hexagonal C grid. We show that the staggered discretizations
(triangular B and C grids and hexagonal C grid) maintain spurious
instability branches on both baroclinic and symmetric axes. These
branches are at least partly induced by numerical modes of these
discretizations. This is indicated by the loss of Galilean invariance
for these branches. The collocated triangular A grid has a spurious
branch on the baroclinic axis, but no such branches on the symmetric
axis. The spurious instability branches are sensitive to the alignment
between the mean flow and the mesh and to the discretization of the
advection of momentum and scalars.

We demonstrate that these grids have very similar accuracy in
representing the physical branch of baroclinic instability, with C
grids being slightly more accurate than the A and B grids at the
coarsest mesh considered here, similar to quadrilateral meshes. On the
symmetric axis, all staggered grids are contaminated by spurious
branches which penetrate into the region of physical
instabilities. The behavior is similar to that induced by the
Hollingsworth instability, but is, at least partly, related to the
presence of geometrical modes maintained by discretizations on triangular meshes with staggered variable placement.

Moderate biharmonic dissipation of momentum and buoyancy is found to
be sufficient to suppress the spurious branches on the baroclinic axis
if meshes are fine enough, but dissipation coefficients, $\mathbb{V}$,
as high as 0.01 m/s are needed when the resolution is
coarsened. However, similar levels of dissipation are also needed on
the quadrilateral A and C grids in this case. The level of dissipation
needed to suppress spurious branches on the symmetric axis requires
additional studies, as spurious branches intersect with physical ones
on triangular and hexagonal staggered grids. We only show that
moderate dissipation is sufficient to suppress spurious instability at
spectral end.

We show that triangular and hexagonal C grids do not reach the
performance of quadrilateral C grid. Similarly, triangular B grid does
not reach the performance of quadrilateral B grid. Unexpectedly
enough, the triangular A grid is least problematic and is not any
worse than other grids considered here in simulating instabilities in
the framework of the Eady problem.

The main recommendation of this study is that a careful choice of
dissipation is needed to suppress spurious behavior for the staggered
discretizations considered here, especially if instabilities along the
symmetric axis are of interest.

The supplementary material can be accessed at \cite{Maass26}.

\section*{Acknowledgment}
This work is a contribution to projects M3
and S2 of the Collaborative Research Centre TRR181 “Energy Transfer in Atmosphere and Ocean” funded by the
Deutsche Forschungsgemeinschaft (DFG, German Research Foundation)—Projektnummer 274762653.

\appendix
\section{Vertical Discretization}
\label{sec:appendixa}
Details of the vertical discretization with Lorenz staggering are not provided in the main text and the reader is referred to the discussion by Barham et al. \cite{Barham2018}.
The vertical discretization transfers the linearized operators between horizontal spaces to operators between tensor products of the respective horizontal space with a vertical space, $V$.
In this study a linearized free surface is applied at the top which simplifies the expression of the pressure in terms of the buoyancy; in wavenumber space the relation may be formulated in terms of a matrix $\underline{\mathsf{P}}$ representing an endomorphism on $V$: $\underline{\Hat{p}} = (\mathsf{I}\otimes\underline{\mathsf{P}})\underline{\Hat{b}}$.
Here $\mathsf{I}$ denotes the identity on the respective horizontal space (one-dimensional complex space on triangular B grids and two-dimensional complex space on triangular C grids).
Integration of the discretized hydrostatic balance \eqref{eq:hydrostatic-eq} applying the homogeneous Dirichlet boundary condition \eqref{eq:top-bc} at the top yields the following expression for the operator $\underline{\mathsf{P}}$:
\begin{eqnarray}\label{eq:Pop}
    \underline{\mathsf{P}} &=& 
    \begin{bmatrix}
        1 & \cdots & \cdots & 1\\
         & \ddots      &        &  \vdots \\
        &       & 1 & 1 \\
         &   &  & 1\\
         &  &  & 
    \end{bmatrix}
    \left(-\Delta_z \right)
    \frac{1}{2}\begin{bmatrix}
        1 & 1 &        &         & \\
          & 1 & 1      &         & \\
          &   & \ddots & \ddots  & \\
          &   &        & 1       & 1
    \end{bmatrix}\nonumber\\
    &=& 
    -\frac{\Delta_z}{2}\begin{bmatrix}
        1 & 2 & \cdots & \cdots  & 2 & 1\\
          & 1 & 2      &  \cdots & 2 & 1 \\
          &   & \ddots & \ddots  & \vdots & \vdots\\
          &   &        & 1       & 2 & 1\\
          &   &        &         & 1 & 1\\
          &   &        &         &   &
    \end{bmatrix}
\end{eqnarray}

For completeness also the expression between the horizontal velocities and the vertical velocities is shortly repeated.
Due to the Lorenz staggering the vertical velocity field must be averaged to mid-level layers.
Following the notation in Barham et al. \cite{Barham2018} the operator is denoted by $\mathrm{a_z^+}$.
For example, in wavenumber space the contributions of vertical advection of the background horizontal velocities, $U_z\mathsf{A^{(x)}}\Hat{w}$ and $V_z\mathsf{A^{(y)}}\Hat{w}$, are replaced by the following fully discretized versions: $(\mathsf{A^{(x)}}\otimes\underline{U_z})(\mathsf{I}\otimes\mathsf{a_z^+)\underline{\Hat{w}}}$ and $(\mathsf{A^{(y)}}\otimes\underline{V_z})(\mathsf{I}\otimes\mathsf{a_z^+)\underline{\Hat{w}}}$.
As above $\mathsf{I}$ denotes the identity on the appropriate horizontal space.
In wavenumber space the relation between vertical and horizontal velocities may be formulated in terms of the matrix $\underline{\mathsf{W}}$: $(\mathsf{I}\otimes\mathsf{a_z^+})\underline{\Hat{w}} = \underline{\mathsf{W}}\underline{\mathbf{\Hat{u}}}$.
Integration of the discretized continuity equation \eqref{eq:continuity-eq} applying the homogenous Dirichlet boundary condition \eqref{eq:bottom-bc} at the bottom yields the following expression for the matrix $\underline{\mathsf{W}}$:
\begin{eqnarray}\label{eq:Wop}
    \underline{\mathsf{W}} &=& 
    -\mathsf{D} \otimes 
    \frac{\Delta_z}{2}\begin{bmatrix}
        1 & 1 &        &         & \\
          & 1 & 1      &         & \\
          &   & \ddots & \ddots  & \\
          &   &        & 1       & 1
    \end{bmatrix}
    \begin{bmatrix}
        0 & \cdots & \cdots & 0\\
        1 & 0      &        &  \vdots \\
        1 & 1      & \ddots & \vdots \\
        \vdots &   & \ddots & 0\\
        1 & \cdots & \cdots & 1
    \end{bmatrix}\nonumber\\
    &=&
    -\mathsf{D} \otimes 
    \frac{\Delta_z}{2}\begin{bmatrix}
        1 &          &         & \\
        2 & 1        &         & \\
        \vdots    & \ddots & \ddots  & \\
        2 &   \cdots        & 2       & 1
    \end{bmatrix}.
\end{eqnarray}

\section{Geometry of Triangular Meshes}
\label{sec:appendixb}
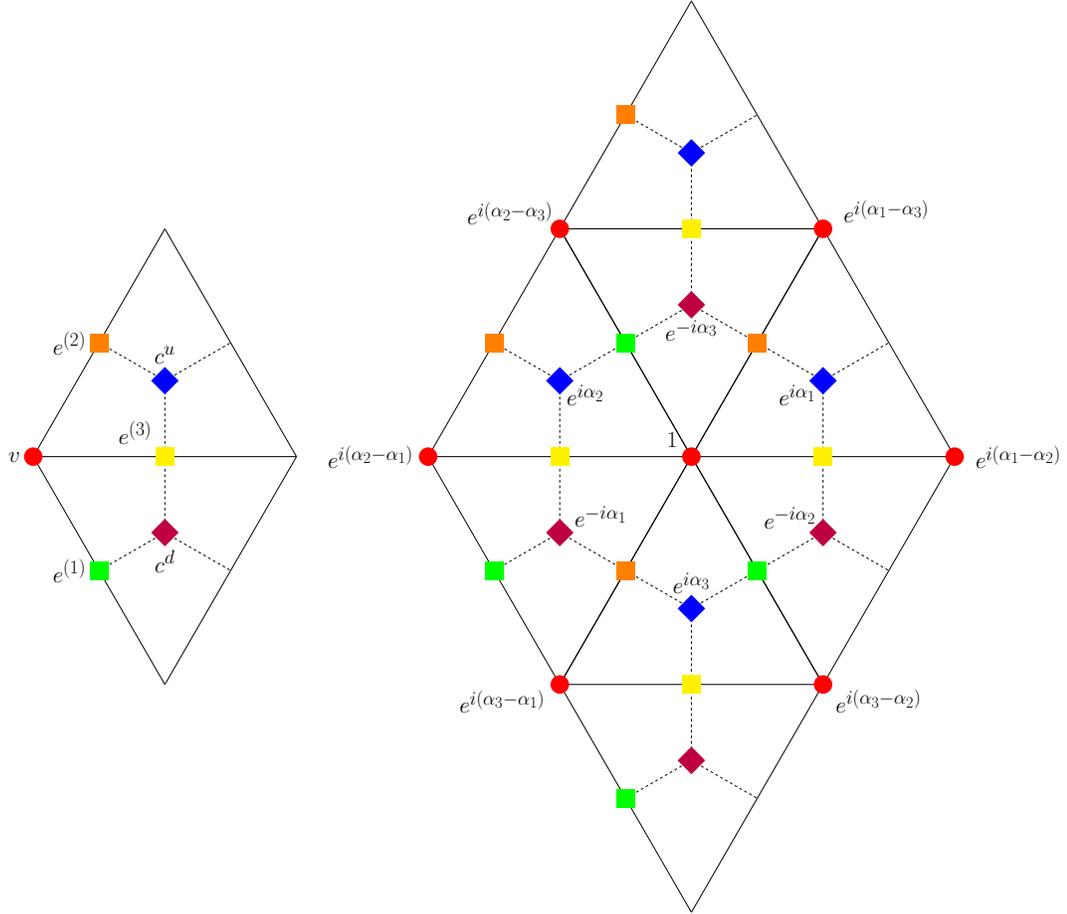
\begin{figure*}
    \centering
    \scalebox{0.35}{
    \begin{tikzpicture}
    \foreach \x/\y in {-15/0}
    {
        \draw (\x+0,{\y+0}) -- ({\x+10},{\y+0});
        \draw ({\x+0},{\y+0}) -- ({\x+5},{\y+sqrt(3)/2*10});
        \draw ({\x+0},{\y+0}) -- ({\x+5},{\y+ -sqrt(3)/2*10});
        \draw ({\x+10},{\y+0}) -- ({\x+5},{\y+ -sqrt(3)/2*10});
        \draw ({\x+10},{\y+0}) -- ({\x+5}, {\y+sqrt(3)/2*10});
        \draw[dashed] ({\x+5},{\y-1/(sqrt(3)*2)*10}) -- ({\x+5},{\y+1/(sqrt(3)*2)*10});
        \draw[dashed] ({\x+5},{\y-1/(sqrt(3)*2)*10}) -- ({\x+2.5},{\y+ -sqrt(3)/4*10});
        \draw[dashed] ({\x+5},{\y+1/(sqrt(3)*2)*10}) -- ({\x+2.5},{\y+ sqrt(3)/4*10});
        \draw[dashed] ({\x+5},{\y-1/(sqrt(3)*2)*10}) -- ({\x+7.5},{\y+ -sqrt(3)/4*10});
        \draw[dashed] ({\x+5},{\y+1/(sqrt(3)*2)*10}) -- ({\x+7.5},{\y+ sqrt(3)/4*10});
    }
    \foreach \x/\y in {-15/0}
    {
        \node[circle, color=red, fill=red, minimum size=20pt, label={[font=\Huge]left:$v$}] (v1) at ({\x+0},{\y+0}) {};

        \node[diamond, color=blue, fill=blue, aspect=1, minimum size=30pt, label={[font=\Huge]above:$c^u$}] (c1) at ({\x+5},{\y+1/(sqrt(3)*2)*10}) {};
        \node[diamond, color=purple, fill=purple, aspect=1, minimum size=30pt, label={[font=\Huge]below:$c^d$}] (c1) at ({\x+5},{\y+ -1/(sqrt(3)*2)*10}) {};
        \node[rectangle, color=orange, fill=orange, minimum height=20pt, minimum width=20pt, label={[font=\Huge]left:$e^{(2)}$}] (e2) at ({\x+2.5}, {\y+sqrt(3)/4*10}) {};

        \node[rectangle, color=green, fill=green, minimum height=20pt, minimum width=20pt, label={[font=\Huge] left:$e^{(1)}$}] (e1) at ({\x+2.5}, {\y+ -sqrt(3)/4*10}) {};

        \node[rectangle, color=yellow, fill=yellow, minimum height=20pt, minimum width=20pt, label={[font=\Huge]above left:$e^{(3)}$}] (e3) at ({\x+5}, {\y+0}) {};
    }
    
    \foreach \x/\y in {0/0, 10/0, 5/{sqrt(3)/2*10}, 5/{-sqrt(3)/2*10}}
    {
        \draw (\x+0,{\y+0}) -- ({\x+10},{\y+0});
        \draw ({\x+0},{\y+0}) -- ({\x+5},{\y+sqrt(3)/2*10});
        \draw ({\x+0},{\y+0}) -- ({\x+5},{\y+ -sqrt(3)/2*10});
        \draw ({\x+10},{\y+0}) -- ({\x+5},{\y+ -sqrt(3)/2*10});
        \draw ({\x+10},{\y+0}) -- ({\x+5}, {\y+sqrt(3)/2*10});
        \draw[dashed] ({\x+5},{\y-1/(sqrt(3)*2)*10}) -- ({\x+5},{\y+1/(sqrt(3)*2)*10});
        \draw[dashed] ({\x+5},{\y-1/(sqrt(3)*2)*10}) -- ({\x+2.5},{\y+ -sqrt(3)/4*10});
        \draw[dashed] ({\x+5},{\y+1/(sqrt(3)*2)*10}) -- ({\x+2.5},{\y+ sqrt(3)/4*10});
        \draw[dashed] ({\x+5},{\y-1/(sqrt(3)*2)*10}) -- ({\x+7.5},{\y+ -sqrt(3)/4*10});
        \draw[dashed] ({\x+5},{\y+1/(sqrt(3)*2)*10}) -- ({\x+7.5},{\y+ sqrt(3)/4*10});
    }
    \foreach \x/\y in {0/0, 10/0, 5/{sqrt(3)/2*10}, 5/{-sqrt(3)/2*10}}
    {
        \node[circle, color=red, fill=red, minimum size=20pt] (v1) at ({\x+0},{\y+0}) {};

        \node[diamond, color=blue, fill=blue, aspect=1, minimum size=30pt] (c1) at ({\x+5},{\y+1/(sqrt(3)*2)*10}) {};
        \node[diamond, color=purple, fill=purple, aspect=1, minimum size=30pt] (c1) at ({\x+5},{\y+ -1/(sqrt(3)*2)*10}) {};
        \node[rectangle, color=orange, fill=orange, minimum height=20pt, minimum width=20pt] (e2) at ({\x+2.5}, {\y+sqrt(3)/4*10}) {};

        \node[rectangle, color=green, fill=green, minimum height=20pt, minimum width=20pt] (e1) at ({\x+2.5}, {\y+ -sqrt(3)/4*10}) {};

        \node[rectangle, color=yellow, fill=yellow, minimum height=20pt, minimum width=20pt] (e3) at ({\x+5}, {\y+0}) {};
    }

    \node[label={[font=\Huge, label distance=3mm]150: $1$}] at (10, 0) {};
    \node[label={[font=\Huge]190: $e^{i\alpha_1}$}] at (15, {1/(sqrt(3)*2)*10}) {};
    \node[label={[font=\Huge]350: $e^{i\alpha_2}$}] at (5, {1/(sqrt(3)*2)*10}) {};
    \node[label={[font=\Huge, label distance=3mm]90: $e^{i\alpha_3}$}] at (10, {-2/(sqrt(3)*2)*10}) {};
    
    \node[label={[font=\Huge, label distance=3mm]10: $e^{-i\alpha_1}$}] at (5, {-1/(sqrt(3)*2)*10}) {};
    \node[label={[font=\Huge]170: $e^{-i\alpha_2}$}] at (15, {-1/(sqrt(3)*2)*10}) {};
    \node[label={[font=\Huge, label distance=3mm]270: $e^{-i\alpha_3}$}] at (10, {2/(sqrt(3)*2)*10}) {};

    \node[label={[font=\Huge, label distance=3mm]190: $e^{i(\alpha_3-\alpha_1)}$}] at (5, {-sqrt(3)/2*10}) {};
    \node[label={[font=\Huge]170: $e^{i(\alpha_2-\alpha_3)}$}] at (5, {sqrt(3)/2*10}) {};
    \node[label={[font=\Huge, label distance=3mm]180: $e^{i(\alpha_2-\alpha_1)}$}] at (0, 0) {};
    \node[circle, color=red, fill=red, minimum size=20pt,label={[font=\Huge, label distance=3mm]10: $e^{i(\alpha_1-\alpha_3)}$}] at (15, {sqrt(3)/2*10}) {};
    \node[circle, color=red, fill=red, minimum size=20pt,label={[font=\Huge]350: $e^{i(\alpha_3-\alpha_2)}$}] at (15, {-sqrt(3)/2*10}) {};
    \node[circle, color=red, fill=red, minimum size=20pt,label={[font=\Huge, label distance=3mm]0: $e^{i(\alpha_1-\alpha_2)}$}] at (20, 0) {};
    
    \end{tikzpicture}
}
    \caption{\textit{Left}: Primitive unit cell of a hexagonal lattice. The red circle is the representative of the hexagonal lattice in the unit cell and located at a vertex, $v$, of a triangle. The unit cell contains two triangle centers, $c^u$ and $c^d$, that form two hexagonal lattices represented by the blue and purple diamond respectively. Moreover, the unit cell contains three edge, $e^{(1)}$, $e^{(2)}$, and $e^{(3)}$, centers that form three hexagonal lattices which are represented by the orange, yellow, and green square respectively. \textit{Right}: Excerpt of the triangular/hexagonal grid centered at a vertex. The hexagon surrounding the vertex is called the \textit{dual control volume} corresponding to the vertex and its corners are the three centers of the upward pointing triangles (blue) and the three centers of the downward pointing triangles (red). The labels of the corners denote the \textit{phase shift} of a plane wave with fixed wavevector, $\boldsymbol{k}$, with respect to the central vertex. The dependence on the wavevector is neglected in the notation.}
    \label{fig:hex-lattice}
\end{figure*}

A regular triangular mesh is the tessellation of the plane by a primitive unit cell (fundamental region) made of two triangles of different orientation.
In the left panel of Figure \ref{fig:hex-lattice} one such choice of primitive unit cell is illustrated.
In the regular triangular mesh the vertices form one hexagonal lattice.
The centroids of the triangles, however, form two different lattices that cannot be superimposed by a translation.
The two lattices are made up of the centroids of the triangles with the different orientations respectively.
In addition, the midpoint of the edges form three distinct lattices.
A numbering of these lattices is introduced in the left panel of Figure \ref{fig:hex-lattice}.

\section{Parameters in the Problem Formulation}

\begin{table*}
    \centering
    \begin{tabular}{l l}
        \toprule
         Name & Interpretation\\
         \midrule
         $g$            & Gravitational acceleration (units: $m s^{-2})$\\
         $f_0$          & Coriolis parameter on a f-plane (units: $s^{-1}$) \\
         $\mathrm{Ri}$  & Richardson number in the background flow\\
         $N$ & Brunt-Väisälä frequency of the background flow (units: $s^{-1}$)\\
         $\mathbb{K}^u$ & Viscosity coefficient ($\mathbb{K}^u = \mathbb{V}^u * a^3$ for biharmonic dissipation)\\
         $\mathbb{K}^b$ & Diffusivity coefficient ($\mathbb{K}^b = \mathbb{V}^b * a^3$ for biharmonic dissipation)\\
         $H$            & Total depth of the fluid (units: $m$)\\
         $N_z$          & Number of vertical layers\\
         $a$            & Lattice constant of the hexagonal lattices (units: $m$)\\
         $\theta_U$     & Azimuth of background flow direction in coordinate system of the grid (units: $\mathrm{rad})$\\
         $\beta$        & Speed of the background flow at mid-level in terms of $\abs{HM^2/f_0}$\\
         \bottomrule
    \end{tabular}
    \caption{Summary of the parameters in the problem formulation}
    \label{tab:parameters}
\end{table*}

\newpage

\bibliographystyle{elsarticle-num}
\bibliography{references}

\end{document}